\documentclass{article}
\usepackage{graphicx,amssymb,amsmath}
\usepackage{amsthm}
\usepackage{xspace,framed}
\usepackage[svgnames]{xcolor}
\usepackage{cleveref}
\usepackage{stfloats}
\usepackage{flushend}

\usepackage{float}
\newtheorem{definition}{Definition}
\newtheorem{theorem}{Theorem}

\newtheorem{lemma}[theorem]{Lemma}

\newcommand{\FPT}{\textrm{\textup{FPT}}\xspace}
\newcommand{\WOH}{\textrm{\textup{W[1]-hard}}\xspace}

\newcommand{\NPH}{\textrm{\textup{NP-hard}}\xspace}
\newcommand{\NPC}{\textrm{\textup{NP-complete}}\xspace}

\newcommand{\name}[1]{\textsc{#1}}
\newcommand{\YES}{\textsc{Yes}}
\newcommand{\NO}{\textsc{No}}

\newcommand{\RRE}{\textsc{Runaway Rectangle Escape Problem}}

\newcommand{\OPT}{\mathcal{OPT}}

\newcommand{\AAA}{{\mathcal A}}

\newcommand{\LL}{{\mathcal L}}

\newcommand{\PP}{{\mathcal P}}

\newcommand{\RR}{{\mathcal R}}
\newcommand{\SSS}{{\mathcal S}}
\newcommand{\TT}{{\mathcal T}}

\newcommand{\ran}{runaway number}
\newcommand{\RE}{\name{Rectangle Escape Problem}}

\usepackage{graphicx}
\usepackage{tikz}
\usetikzlibrary{decorations.shapes}
\makeatletter
\def\grd@save@target#1{%
  \def\grd@target{#1}}
\def\grd@save@start#1{%
  \def\grd@start{#1}}
\tikzset{
  grid with coordinates/.style={
    to path={%
      \pgfextra{%
        \edef\grd@@target{(\tikztotarget)}%
        \tikz@scan@one@point\grd@save@target\grd@@target\relax
        \edef\grd@@start{(\tikztostart)}%
        \tikz@scan@one@point\grd@save@start\grd@@start\relax
        \draw[minor help lines] (\tikztostart) grid (\tikztotarget);
        \draw[major help lines] (\tikztostart) grid (\tikztotarget);
        \grd@start
        \pgfmathsetmacro{\grd@xa}{\the\pgf@x/1cm}
        \pgfmathsetmacro{\grd@ya}{\the\pgf@y/1cm}
        \grd@target
        \pgfmathsetmacro{\grd@xb}{\the\pgf@x/1cm}
        \pgfmathsetmacro{\grd@yb}{\the\pgf@y/1cm}
        \pgfmathsetmacro{\grd@xc}{\grd@xa + \pgfkeysvalueof{/tikz/grid with coordinates/major step}}
        \pgfmathsetmacro{\grd@yc}{\grd@ya + \pgfkeysvalueof{/tikz/grid with coordinates/major step}}
      }
    }
  },
  minor help lines/.style={
    help lines,
    step=\pgfkeysvalueof{/tikz/grid with coordinates/minor step}
  },
  major help lines/.style={
    help lines,
    line width=\pgfkeysvalueof{/tikz/grid with coordinates/major line width},
    step=\pgfkeysvalueof{/tikz/grid with coordinates/major step},
  },
  grid with coordinates/.cd,
  minor step/.initial=0.5,
  major step/.initial=1.5,
  major line width/.initial=0.5pt,
}
\makeatother

\title{The Runaway Rectangle Escape Problem}

\newcommand*\samethanks[1][\value{footnote}]{\footnotemark[#1]}

\author{Aniket Basu Roy\thanks{Department of Computer Science and Automation, Indian Institute of Science, {\tt  aniket.basu$|$gsat@csa.iisc.ernet.in}} \and Sathish Govindarajan\samethanks{} \and Anil Maheshwari\thanks{School of Computer Science, Carleton University, {\tt anil@scs.carleton.ca}} \and Neeldhara Misra\thanks{Indian Institute of Technology, Gandhinagar, {\tt neeldhara.m@iitgn.ac.in }} \and Subhas C Nandy\thanks{Advanced Computing and Microelectronics Unit, Indian Statistical Institute, Kolkata, {\tt nandysc@isical.ac.in}} \and Shreyas Shetty\thanks{Department of Computer Science and Engineering, Indian Institute of Technology, Madras, {\tt shshett@cse.iitm.ac.in}}}

\index{Basu Roy, Aniket}
\index{Govindrajan, Sathish}
\index{Maheshwari, Anil}
\index{Misra, Neeldhara}
\index{Nandy, Subhas C}
\index{Shetty, Shreyas}

\newcommand\blfootnote[1]{%
  \begingroup
  \renewcommand\thefootnote{}\footnote{#1}%
  \addtocounter{footnote}{-1}%
  \endgroup
}

\begin{document}

\maketitle

\begin{abstract}

Motivated by the applications of routing in PCB buses, the \RE{} was recently introduced and studied.
In this problem, we are given a set of rectangles $\SSS$ in a rectangular region $R$, and we would like to extend these rectangles to one of the four sides of $R$. Define the density of a point $p$ in $R$ as the number of extended rectangles that contain $p$. The question is then to find an extension with the smallest maximum density. 

We consider the problem of maximizing the number of rectangles that can be extended when the maximum density allowed is at most $d$. It is known that this problem is polynomially solvable for $d = 1$, and NP-hard for any $d \geq 2$. We consider approximation and exact algorithms for fixed values of $d$. We also show that a very special case of this problem, when all the rectangles are unit squares from a grid, continues to be \NPH{} for $d = 2$. 
\end{abstract}

\blfootnote{
A preliminary version of this work appeared in the Proceedings of the $26^{\textrm{th}}$ Canadian Conference on Computational Geometry, 2014 \cite{DBLP:conf/cccg/RoyGMS14}.
}

\section{Introduction}

The \RE{} was introduced in~\cite{MKWY11}, and was further explored in~\cite{AEYZ13}. In its original formulation, the problem is the following. Let $\SSS$ be a set of rectangles in a rectangular region $R$. The goal is to extend these rectangles to one of the four sides of $R$ while ensuring that the maximum number of overlaps is minimized. In particular, define the density of a point $p$ in $R$ as the number of extended rectangles that contain $p$. The question is then to find an extension with the smallest maximum density. 

The problem finds its motivation in a closely related escape routing problem on the bus levels in PCBs. A detailed exposition of how the formulation above captures the essence of the bus-routing problem is provided in~\cite{AEYZ13}.

It turns out, by the combined results in~\cite{MKWY11,AEYZ13}, that this question is intractable --- indeed, it is \NPH{} to determine if all rectangles can be extended with density at most $d$ for any \emph{fixed} $d \geq 2$ (the result was known for $d \geq 3$ in~\cite{MKWY11} and was established for $d = 2$ in~\cite{AEYZ13}), even when the given set of rectangles are disjoint to begin with. However, the case when $d=1$ is solvable in polynomial time --- the first proposed algorithm from~\cite{KMYW10} had a running time of $O(n^6)$. Subsequently, Assadi et al.\ demonstrate a dynamic programming approach with an improved running time of $O(n^4)$ in~\cite{AEYZ13}. The recursive formulations in the DP involve finding the maximum number of rectangles that can be routed in a given subset of directions while being completely disjoint in their extended state. For the problem of optimizing the density, a factor-$4$ approximation is known in general (by standard rounding techniques), and a PTAS can be obtained on the assumption that the optimal density is high. We refer the readers to~\cite{AEYZ13} for details. 


In certain scenarios, the density of any point $p$ in $R$ cannot exceed a threshold value $d$, which is fixed by practical considerations. 
Here, the natural question is to maximize the number of rectangles that can be extended, subject to this fixed density $d$. We call this problem, the $d$-\RRE{}. Note that this problem is complementary to the original \RE{}. The \RE{} asks to extend all the rectangles minimizing the density. The $d$-\RRE{} asks to extend the maximum number of rectangles subject to density at most $d$, for any input parameter $d$.

This problem turns out to have strong connections to the classical $k$-fold packing problem, which is a generalization of the independent set problem. Precisely, the $d$-\RRE{} in $\mathbb{R}^m$ with density parameter $d$ is related to the $(d-1)$-fold packing of boxes in $\mathbb{R}^{m-1}$ to achieve approximate solutions. 

The $k$-fold packing problem for geometric objects in $\mathbb{R}^m$ is defined as follows. Given a set of objects $\RR$ in $\mathbb{R}^m$ and an integer parameter $k$, a $k$-fold packing $\RR'\subseteq\RR$ is a subset of the set of input objects such that for every point $x\in \mathbb{R}^m$, $|\{R \mid x\in R, \forall R \in \RR'\}| \le k$, i.e., the number of objects in $\RR'$  containing $x$ is at most $k$ for every $x\in \mathbb{R}^m$ \cite[Chap 2]{toth2004handbook}. 
Note that for $k=1$ it is an independent set of the objects. The problem is to maximize the size of the $k$-fold packing. Ene et al.\ \cite{Ene2011a} have studied approximation algorithms for this problem in geometric context. For our purposes, the objects are intervals and rectangles. The problem is polynomial time solvable for intervals \cite{faigle1995note} whereas iteratively using the LP based algorithm for solving the maximum independent set of rectangles, due to Chalermsook and Chuzhoy \cite{DBLP:conf/soda/ChalermsookC09}, one can get an $O(\log\log n)$ approximation factor.

Th $d$-\RRE{} is clearly \NPH{} for $d \geq 2$, since it is \NPH{} to determine whether all rectangles can be extended, i.e., is the optimum for this problem equal to $n$? We explore this problem from the point of view of approximation and fixed-parameter tractability. We also give a randomized algorithm that outputs an $O(1-\epsilon)$-approximate solution with appropriate restrictions on the density.
On the approximation front, we show that if the rectangles are disjoint, then we have a $4(1 + 1/(d-1))$-approximation for the problem, and in general, we have a $(4d)$-approximation. We further extend the ideas to boxes, $3$-dimensional rectangles, where for disjoint boxes we have an $O(d \log\log n)$-approximation, and in general, we have an $O(d(\log\log n)^3)$-approximation. 

For the disjoint setting, we use the $k$-fold packing of the projections of rectangles (boxes), i.e., intervals (rectangles) to get the approximation algorithm. Moreover, we show that the implications are also in the other direction, i.e., better approximations for the $d$-\RRE{} where the boxes are disjoint implies better approximation for the $(d-1)$-fold packing of rectangles problem. 
We will describe this in more detail at the end of section \ref{subsec:disjoint}.



The decision version of the $d$-\RRE{} may be informally stated as follows: are there at least $k$ rectangles that can be extended with density at most $d$? There are two natural parameters for this problem; namely $k$, the number of rectangles that we wish to extend, and $d$, the maximum density that is allowed. Since the problem is \NPC{} even for constant values of $d$, we do not expect this problem to be fixed-parameter tractable parameterized by $d$ alone. On the other hand, we show that when parameterized by $k$, for fixed $d$, the problem is indeed fixed-parameter tractable, as long as the input rectangles have density at most $(d-1)$. 

We also consider the following closely related question: can we extend at least $p$ non-boundary rectangles horizontally (i.e., towards the right or left), and at least $q$ non-boundary rectangles vertically (i.e., towards the top or bottom) without violating the density constraints? A non-boundary rectangle is one that doesn't share an edge with the boundary of $R$. It is natural to consider only non-boundary rectangles in our demand for extension, since the ones on the boundary, without loss of generality, can be ``extended'' to the boundary that they are on. We show that this problem is \WOH{}, which implies that a fixed-parameter tractable algorithm does not exist unless the \name{Exponential Time Hypothesis} fails. 

Finally, we consider a special case of \RE{} when all the rectangles are unit squares aligned to an underlying grid. For this problem when $d=2$, we show a non-trivial reduction from a variant of \name{Not-All-Equals SAT}, establishing NP-hardness. We also show that this problem enjoys a 2-factor approximation algorithm.

\section{Preliminaries}

Let $\SSS$ be a set of rectangles in a rectangular region $R$. For $\TT \subseteq \SSS$, let $\Gamma(\SSS,\TT)$ be obtained from $\SSS$ by extending each rectangle in $\TT$ to one of the four borders of $R$. We call $\Gamma(\SSS,\TT)$ the \emph{extended configuration of $\SSS$ with respect to $\TT$}. Further, we say that $\Gamma(\SSS,\TT)$ has \emph{density at most $d$} if every point in $R$ is contained in at most $d$ rectangles in the extended configuration.  We will refer the maximum density over all points in $R$, before extending the rectangles, to be the \textit{input density} of $\SSS$.

For a fixed $d$, the size of the largest subset $\TT$ for which the density of $\Gamma(\SSS,\TT) \leq d$ is called the \emph{\ran{} of $\SSS$ with respect to $d$}, which we denote by $\rho(\SSS,d)$. We study the following decision version of the $d$-\RRE{}:

\begin{framed}
\noindent \name{{\sc $d$-\RRE{} --- Decision version}}

\noindent Input: A set of $n$ rectangles $\SSS$ in a rectangular region $R$, and an integer $k$.\\
Question: Is $\rho(\SSS,d) \geq k$?
\end{framed}

The exact algorithms are considered in the framework of parameterized complexity. We only introduce the terminology that we use in this work, the reader is referred to the books~\cite{niedermeier2006invitation,flum2006parameterized,downey1999parameterized} for a comprehensive exposition. A parameterized problem $\Pi$ is a subset of $\Gamma^{*}\times\mathbb{N}$, where $\Gamma$ is a finite alphabet. An instance of a
parameterized problem is a tuple $(x,k)$, where $x$ is the input string and $k$ is called the
parameter. A central notion in parameterized complexity is {\em
  fixed-parameter tractability (FPT)} which means, for a given
instance $(x,k)$, decidability is in time $f(k)\cdot p(|x|)$, where
$f$ is a computable function of $k$ and $p$ is a polynomial in the
input size $|x|$. We now define the notion of parameterized reduction.

%
%

\begin{definition}
Let $A,B$ be parameterized problems.  We say that $A$ is
(uniformly many:1) {\bf \em fpt-reducible} to $B$ if there exist functions 
$f,g:\mathbb{N}\rightarrow \mathbb{N}$, a constant $\alpha \in \mathbb{N}$ and 
 an algorithm $\Phi$ which transforms an instance $(x,k)$ of $A$ into an instance $(x',g(k))$ of $B$ 
in time $f(k) |x|^{\alpha}$ so that $(x,k) \in A$ if and only if $(x',g(k)) \in B$.
\end{definition}


A parameterized problem is considered unlikely to be fixed-parameter tractable
if it is $W[i]$-hard for some $i\geq 1$. To show that a problem is
$W[1]$-hard, it is enough to give a parameterized reduction from a known
$W[1]$-hard problem. It is well known that the parameterized
version of the {\textsc{Maximum Clique Problem}} is $W[1]$-hard. In particular,
we use the following popular variant of the problem~\cite{Fellows200953}:

\begin{framed}
\noindent \name{{\sc Multi-Colored Clique}}

\noindent Input: A graph $G$ whose vertex set is partitioned into $k$ parts, $V_1 \uplus \cdots \uplus V_k$.\\
Question: Is there a subset $S$ of vertices such that $G[S]$ is a clique and $|S \cap V_i| = 1$ for all $1 \leq i \leq k$?\\
Parameter: $k$.
\end{framed}

\section{Approximation Algorithms}
In this section, we consider the optimization version of the problem, where the goal is to extend the maximum number of rectangles with density at most $d$. We devise approximation algorithms for rectangles in the plane and also extend the idea for boxes, i.e., rectangles in $3$-dimensional space. Thereafter, we assume that the rectangles are disjoint in the input instance, and show that better approximations can be reached under this assumption.

In order to approximate the optimum solution we will focus our attention to a particular direction or axis and look into the problem restricted to that direction or axis $\lambda\in\Lambda$ where $\Lambda$ is the set of directions (axes), i.e., the rectangles are allowed to extend only in the direction (along the axis) $\lambda$. 
We denote the optimum solution by $\OPT$ and when the problem is restricted to $\lambda$, then its optimum solution is denoted by $\OPT_\lambda$.
We claim the following statement.
\begin{lemma}
\label{pigeon}
There exists a direction (axis) $\lambda^*\in\Lambda$ such that \[|\OPT_{\lambda^*}|\ge\frac{|\OPT|}{|\Lambda|}\]
\end{lemma}

\proof
We partition the optimum solution into sets $\mathcal{S}_\lambda$ such that in each set every rectangle is extended in the direction (axis) $\lambda$.
\[\OPT=\displaystyle\biguplus_{\lambda\in\Lambda}\mathcal{S}_{\lambda}\]
As the maximum is at least the average there exists a direction (axis) $\lambda^*$ such that $|\mathcal{S}_{\lambda^*}|\ge |\OPT|/|\Lambda|$.
For every $\lambda$, $|\OPT_\lambda| \ge |\mathcal{S}_{\lambda}|$ because $\mathcal{S}_{\lambda}$ is a feasible solution to the problem restricted to the direction (axis) $\lambda$. 
Together with the previous inequality, the claim follows.
\qed
\subsection{Arbitrary Family of Rectangles}
Let $(R,\SSS,d)$ be an instance of $d$-\RRE. Let $\OPT$ denote an optimum solution. Note that at least half the rectangles in $\OPT$ are pushed either horizontally or vertically. So we consider the following problem: given the rectangles $\SSS$, what is the largest number of rectangles, $\OPT_x$, that can be extended vertically with density at most $d$? We show that this can be approximated to within a factor of $2d$, and repeating the argument along the horizontal direction, and reporting the best of both solutions leads us to a $(4d)$-approximation overall. 

We remark that in~\cite{AEYZ13}, the problem of determining if at least $k$ rectangles can be extended vertically with density one (that is, with no overlapping rectangles) is shown to be polynomially solvable. They use a natural greedy strategy: consider the rectangles in the order of decreasing $y$-coordinates of the bottom edges. Let this order be $R_1, \ldots, R_n$. For $1 \leq i \leq n$, we try to extend $R_i$ upwards if this causes no conflicts, else we attempt to extend it downwards. If $R_i$ is blocked in both directions, we choose not to extend it and move to the next rectangle on the list. However, this strategy does not work as-is, for instance, when $d=2$. 

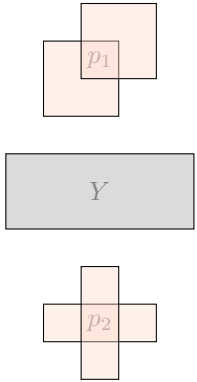
\begin{figure}
\begin{center}
\begin{tikzpicture}[scale=.5, fill opacity=.4, draw opacity=1]
\begin{scope}[xshift=3cm]
\draw [fill=Gray!70] (0,4) rectangle (5,6) node[pos=.5] {$Y$};
\draw [fill=OrangeRed!20] (1,7) rectangle (3,9) node[pos=.75] {$p_1$};
\draw [fill=OrangeRed!20] (2,8) rectangle (4,10);
\draw [fill=OrangeRed!20] (1,1) rectangle (4,2) node[pos=.5] {$p_2$};
\draw [fill=OrangeRed!20] (2,0) rectangle (3,3);
\end{scope}
\end{tikzpicture}
\end{center}
\caption{The notion of stuck rectangles.}
\label{fig:stuck}
\end{figure}

As a preprocessing step, we will first forbid some rectangles from consideration. For a rectangle $Y$ whose $x$-projection is given by the interval $(a,b)$, let us denote by ${\mathcal B}(Y)$ the region between the lines $x = a$ and $x = b$. Call a rectangle $Y$ \emph{stuck} if there are points $p_1,p_2$ of density $d$ (in the input configuration) contained in ${\mathcal B}(Y)$, with $p_1$ above $Y$ and $p_2$ below $Y$, see figure \ref{fig:stuck}. Note that the set of stuck rectangles do not participate in any solution. We now turn our attention to the remaining rectangles, which we refer to as ``good'' rectangles. 

We begin by considering the projections of the good rectangles on the $x$-axis. 
Compute a maximum independent set $\AAA$ among these intervals. Return the rectangles corresponding to $\AAA$. As the returned rectangles are not `stuck' and they do not overlap with each other, they indeed form a feasible solution.


\paragraph*{Analysis.} Note that for intervals a maximum independent set can be computed in polynomial time. Moreover, its size is same as that of the minimum sized piercing set\footnote{A point set in the real line is called a piercing set of the set of intervals if every interval contains at least one point from the set.} as stated by Tibor Gallai in an unpublished work \cite{DBLP:journals/dm/GyarfasL85}. Next consider the projection of the rectangles in $OPT_x$. Observe that the size of the maximum clique in this set of intervals is at most $2d$. If not, then there exists a point on the $x$-axis with density more than $2d$. This means that there are more than $d$ rectangles extended, either upwards or downwards, that are stabbed by the vertical line passing through this point having density more than $2d$.
Hence, $|OPT_x| \le 2d|\AAA|$. Thus, as claimed, the problem restricted to a direction can be approximated within a factor of $2d$. A similar argument holds for the problem of extending rectangles in the horizontal direction, and the better of the two solutions is a $(4d)$-approximation overall.



\begin{theorem}
There exists a polynomial time $4d$-approximation algorithm for the \textsc{$d$-Runaway Rectangle Escape Problem}.
\end{theorem}

\subsubsection*{Boxes}
Here, in addition to the $2$ axes (i.e., horizontal and vertical) we also have the lateral axis ($z$-axis). We consider the following restricted problem: what is the largest number of rectangles that can be extended along a fixed axis with density at most $d$? 
Lemma \ref{pigeon} implies that there exists an axis along which at least one-third of the rectangles in the optimum solution, $\OPT$, are extended. We solve the $3$ restricted problems and return the best of the $3$ solutions.


Hereafter, we look into the restricted problem where boxes are allowed to extend only along the $z$-axis. We will return a solution that is approximate to the optimum within a factor of $O(d(\log\log n)^3)$. 
The approximation ratio suffers for boxes due to the approximation factors involved in computing the maximum independent set of rectangles and the upper bound on the duality gap between the piercing and independent set problems for rectangles.

We project every box in each of the three orthogonal planes. Without loss of generality, we stick to the $xy$\nobreakdash-plane. 
Similar to the $2$-dimensional case, we call a box stuck if extending the box alone in either of the directions in the $z$-axis violates the density constraint.
We will forbid the stuck boxes from consideration. 

Now, we have a set of rectangles $\RR$ in the $xy$\nobreakdash-plane, after removing the stuck boxes. We compute an independent set $\mathcal{A}$ with the best known polynomial time approximation algorithm. We return the boxes corresponding to the rectangles in the computed independent solution. The solution is feasible as these boxes are disjoint and good.

\paragraph*{Analysis.} Let $\nu$ be the size of the maximum independent set in $\RR$. Using the algorithm by Chalermsook and Chuzhoy \cite{DBLP:conf/soda/ChalermsookC09}, the size of $\mathcal{A}$ is at least $\Omega(\nu/\log\log|\RR|)$. Thus, the approximation factor is $O(\log\log n)$ as $|\RR|\le n$, where $n$ is the number of input boxes.
Now consider the minimum sized piercing set\footnote{A point set in plane is called a piercing set of $\RR$ if every rectangle contains at least one point from the set.} of $\RR$ where the size of such a set is denoted by $\tau$. In fact, also consider the set of straight lines $\LL$ parallel to the $z$-axis passing through the piercing points in the $xy$\nobreakdash-plane.
We will call the straight line $L_p$ that is parallel to the $z$-axis passing through the piercing point $p$ in the $xy$\nobreakdash-plane.
So, all the good boxes are stabbed by some line $L_p$ because their projections on the $xy$\nobreakdash-plane are pierced by some point $p$.
Now, consider the projection on the $xy$\nobreakdash-plane of the optimum solution, $\OPT_z$, where boxes are restricted to extend only along the $z$-axis. Observe that the maximum clique number of these rectangles can be at most $2d$. If not then there exists a point $q$ in the $xy$\nobreakdash-plane that is covered by at least $2d+1$ rectangles, which means
$L_q$ witnesses density of at least $d+1$ in either positive or negative direction of the $z$-axis. As the minimum piercing number is $\tau$ and the maximum clique size is $2d$ therefore, the number of rectangles (projections of boxes in $\OPT_z$) is at most $2d\tau$.


Correa et al.~\cite{DBLP:dblp_conf/latin/CorreaFS14}, showed an upper bound of $O(\nu(\log\log\nu)^2)$ for $\tau$. Thus, $|\OPT_z|\le O(2d\nu(\log\log\nu)^2)$. Therefore, the approximation factor of the restricted problem is the following.
$$\frac{|\mathcal{A}|}{|\OPT_z|} \ge \frac{\Omega(\nu/\log\log n)}{O(d\nu(\log\log\nu)^2)}$$
$$|\mathcal{A}| \ge \Omega(\frac{|\OPT_z|}{d(\log\log n)^3})$$
As $\nu\le n$, the last inequality holds. Also, returning the best of the $3$ solutions along the $3$ axes will incur an approximation factor of 3, which will not affect the overall approximation factor asymptotically. Hence, there is an $O(d(\log\log n)^3)$-approximation algorithm for computing the \textsc{$d$-Runaway Box Escape Problem}.

\begin{theorem}
There exists a polynomial time $O(d(\log\log n)^3)$-approximation algorithm for the \textsc{Disjoint $d$-Runaway Box Escape Problem}.
\end{theorem}

\subsection{Disjoint Family of Rectangles} 

We consider the $d$-\RRE{} when the rectangles are disjoint, i.e., the input density is at most unity. We recall that this problem remains \NPC{} due to the reduction in~\cite{AEYZ13} for all $d \geq 2$. For this case, we obtain a $4(1+1/(d-1))$-approximation algorithm. Let $\OPT$ be the optimum solution for the given instance of the \textsc{Disjoint} $d$-\RRE{}.
We consider the four directions $\{x_+,x_-,y_+,y_-\}$ and restrict the problem in each one of them.
From Lemma~\ref{pigeon} we know that there exists a direction $\lambda^*\in\{x_+,x_-,y_+,y_-\}$ such that the optimum solution size to the problem restricted to the direction $\lambda^*$,  $|\OPT_{\lambda^*}|\ge|\OPT|/4$.

Without loss of generality, let us assume that $\lambda^*=y_+$. Let $\mathcal{I}$ denote the projections of the input rectangles on the $x$-axis. 
It is well known that an optimum $d$-fold packing for a system of intervals on the real line can be obtained in polynomial time~\cite{faigle1995note}.  
 
Consider an optimum $(d-1)$-fold packing, $\AAA$, of $\mathcal{I}$.
We observe it gives us a feasible solution to the problem restricted to a given direction.
\begin{lemma}
The rectangles corresponding to the $(d-1)$-fold packing, $\AAA$, of $\mathcal{I}$ is a feasible solution to the \textsc{Disjoint} $d$-\RRE{} restricted to extend to the upward (downward) direction.
\end{lemma}
Since the input rectangles are disjoint, the upward extensions of the rectangles corresponding to any $(d-1)$-fold packing will ensure the maximum density to be at most $d$. Let $\nu_{d-1}$ denote the size of $\AAA$. The approximation algorithm outputs the rectangles corresponding to intervals in $\AAA$.

\paragraph*{Analysis.}We note that a $d$-fold packing may not be a feasible solution when the corresponding rectangles are extended upwards. This is because we may have an input rectangle positioned such that it intercepts a density $d$ region from the extension, causing the overall density to ``spill over'' to $(d+1)$. On the other hand, since a $d$-fold packing on the interval projection can be obtained from an upward extension of density at most $d$, we have:

$$\nu_{d-1} \le |\OPT_\lambda| \le \nu_{d}$$

Now, we use the fact that any $d$-fold packing of intervals is a disjoint union of $d$ independent sets of intervals (see, for example,~\cite{faigle1995note}). Let $\mathcal{C}_1, \mathcal{C}_2, ..., \mathcal{C}_d$ be the independent sets in an optimum $d$-fold packing of $\mathcal{I}$, where we index them in non-increasing order of their sizes. Let $t_i=|\mathcal{C}_i|$ and also $t_i\ge t_{i+1}$ for $1\le i<d$. Thus, $\nu_d = \sum_{1\le i\le d} t_i$ and $\nu_{d-1} \ge \nu_d - t_d$, since removing an independent set from a $d$-fold packing yields a feasible $(d-1)$-fold packing. Now, by an averaging argument, we have $t_d\le \nu_d/d$. Thus, $\nu_{d-1}\ge \nu_d(1-1/d)$. Hence, the following holds. 
$$|\AAA| \ge |\OPT_\lambda|(1-1/d)$$

Returning the best among the $4$ solutions, each corresponding to a direction, will incur another factor of $4$. Thus, the overall approximation factor is $4(1+\frac{1}{d-1})$.

\begin{theorem}
There exists a polynomial time $4(1+\frac{1}{d-1})$-approximation algorithm for the \textsc{Disjoint $d$-Runaway Rectangle Escape Problem}.
\end{theorem}

\subsubsection*{Disjoint Boxes}
\label{subsec:disjoint}
Consider a family of non-overlapping boxes. The problem is to maximize the number of boxes that can be extended without violating the given density constraint. Let $\OPT$ be the optimum solution. 

Now, we pose the same problem but restricting ourselves to one of the $6$ directions $\{x_+,x_-,y_+,y_-,z_+,z_-\}$ each time and choose the best of the $6$ solutions. Let the optimum solution when rectangles are allowed to escape only in a given direction $\lambda$ be $\OPT_\lambda$. From Lemma~\ref{pigeon} there exists a direction $\lambda^*$ such that $|\OPT_{\lambda^*}|\ge\frac{|\OPT|}{6}$. Thus for every direction $\lambda$ we would consider approximating $\OPT_\lambda$.

Now we project the boxes on the plane in some direction $\lambda$ to obtain a family of rectangles $\RR$. We compute a $(d-1)$-fold packing $\AAA$ \cite{Ene2011a}, and return the boxes corresponding to it.
 
\paragraph*{Analysis.} We note again that a $d$-fold packing may not be a feasible solution for similar reasons as were in the $2$-dimensional case.
From Ene et al. \cite[Lemma 2.2]{Ene2011a}, we know that repeatedly using an LP-based $\alpha$-approximation algorithm for $k$ times gives a $2\alpha$-approximation algorithm for $k$-fold packing. As Chalermsook and Chuzhoy \cite{DBLP:conf/soda/ChalermsookC09}, gave an $O(\log\log n)$-approximation algorithm for computing maximum independent set of rectangles, we use the same algorithm $d-1$ times to get a $(d-1)$-fold packing, $\AAA$ such that its size is at least $\Omega(\nu_{d-1}/\log\log n)$.

Now we compare $\nu_d$ with $\nu_{d-1}$. Consider the optimum $d$-fold packing $\PP_d$ that is a disjoint union of $t\le O(d^2)$ color classes. The maximum clique size of the intersection graph of a $d$-fold packing of rectangles is $d$, and the chromatic number of such a graph is bounded by the square of the clique number \cite{Asplund1960}. Therefore, the number of color classes is at most $O(d^2)$. The union of the largest $d-1$ such classes is indeed a feasible $(d-1)$-fold packing. Moreover its size is at least $\frac{\nu_d}{t/(d-1)}$. This means,
$$\nu_{d-1} \ge \Omega\left(\frac{\nu_d}{d}\right).$$


Thus, $|\mathcal{A}| \ge \Omega(\nu_d/d \log\log n)$ holds.
Since, $\OPT_\lambda$ is a feasible $d$-fold packing the following approximation holds.
$$|\mathcal{A}| \ge \Omega\left(\frac{|\OPT_\lambda|}{d \log\log n}\right)$$

As before, choosing the best out of the $6$ solutions does not affect the approximation factor asymptotically. Therefore, we have an $O(d \log\log n)$-approximation algorithm for the \textsc{Disjoint $d$-Runaway Box Escape Problem}.

\begin{theorem}
There exists a polynomial time $O(d \log\log n)$-approximation algorithm for the \textsc{Disjoint $d$-Runaway Box Escape Problem}.
\end{theorem}

It is interesting to note that when the boxes are cubes, i.e., equal in all the 3 dimensions, then similar approaches as above will yield $O(d)$-approximation, both for the general and disjoint case. Constant factor approximation algorithms as well as PTAS are known for computing maximum independent set of squares \cite{Chan2011}. For squares the transversal number $\tau \le 4\nu$ where $\nu$ is the independence number \cite{Ahlswede06}. Even independently, $k$-fold packing problem for squares enjoys a PTAS where $k$ is a constant \cite{Aschner2013}.

\paragraph*{Discussion.} Above, we saw how solution to a $(k-1)$-fold packing of rectangles can be used as an approximate solution to the \textsc{Disjoint $k$-Runaway Box Escape Problem}. Now we shall show that given an $\alpha$-approximation algorithm for \textsc{Disjoint $k$-Runaway Box Escape Problem} restricted to one direction there exists an $\alpha k$-approximation algorithm of the $k$-fold packing of rectangles. 
We are given an arbitrary family of rectangles in the plane and we like to compute the maximum $k$-fold packing of the rectangles. We create a disjoint family of boxes in $\mathbb{R}^3$ such that their projection on a fixed direction will give back the original family of rectangles. The claim is that if we have an algorithm that returns an $\alpha$-approximate solution $\AAA$ to the $k$-\RRE{} restricted to the fixed direction, then we can actually get an $\alpha k$-approximate solution to the $k$-fold packing just by returning the rectangles corresponding to $\AAA$. Any feasible solution with density at most $k$ to this runaway problem gives a feasible solution for the $k$-fold packing problem. Let $\OPT$ be the optimum solution to the restricted problem. From the above discussions it is clear that $\nu_{k-1}\le |\OPT| \le \nu_k$ and $\nu_{k-1} \ge \nu_k/k$. Now, $|\AAA|\ge |\OPT|/\alpha$. Thus, $|\AAA| \ge \nu_d/(\alpha k)$. Thus the claim follows.

\section{A Randomized Algorithm}
\label{rand}
In this section we will present a randomized algorithm for the \RRE{}. Our algorithm will follow very closely the algorithm given by Assadi et al. \cite{AEYZ13}. Precisely, we will prove the following theorem.

\begin{theorem}
There is a randomized algorithm for the \RRE{} that outputs a feasible solution whose size is at least $(1-\epsilon)|\OPT|$, with high probability, provided the given density $d\ge \Omega(\ln n/(\epsilon^2\alpha^2))$ and every point has an input density at most $(1-\alpha)d$, where $\alpha$ is a positive fraction independent of $d$.
\end{theorem}

\proof
We will solve the relaxed linear program for the problem and scale the values appropriately before doing a randomized rounding. Then, we shall argue that the solution is feasible and near-optimal with high probability.

Given a family of rectangles $\mathcal{S}$ there are at most $4n^2$ points to capture the constraints of the problem. This is possible by extending the two vertical and two horizontal sides of every rectangle. This would result to a $2n\times 2n$ grid and the set of grid points is denoted by $G$.

For every rectangle $R_i$, we define four variables $r_{i,\uparrow},r_{i,\downarrow},r_{i,\leftarrow}$ and $r_{i,\rightarrow}$ each for the four directions. Next, for every grid point $p$, define $S_p = \{(i,\lambda)\mid p \in R_i^{\lambda}\}$ where $R_i^{\lambda}$ is the extended region outside rectangle $R_i$ after being extended in the direction $\lambda$. We denote $\Lambda$ as the set $\{\uparrow, \downarrow, \leftarrow, \rightarrow\}$. Also, let $d_p$ be the density of $p$ prior to any extension.

The following is the linear program formulation.
\begin{equation*}
\begin{array}{ll@{}ll}
\text{maximize} 		& \displaystyle\sum\limits_{i\in [n], \lambda\in\Lambda} r_{i,\lambda} 	& &\\
\text{subject to} 	& \displaystyle\sum_{(i,\lambda)\in S_p} r_{i,\lambda} + d_p \le d, 		& &\forall p \in G\\
					& \displaystyle\sum_{\lambda\in\Lambda} r_{i,\lambda} \le 1, 				& &\forall i \in [n]\\
					& r_{i,\lambda} \ge 0													& &\forall i \in [n],\, \forall \lambda \in \Lambda
\end{array}
\end{equation*}

Next for every rectangle $R_i$, we do the following random experiment. We extend $R_i$ in at most $1$ of the $4$ directions. The probability that it is extended in the direction $\lambda$ is $(1-\epsilon)r_{i,\lambda}$ and the probability that it is not extended is $\epsilon$ for $\epsilon\in(0,1/2)$. Thus there may be rectangles not extended after this random experiment.

We claim that the solution returned is feasible with high probability. For every rectangle $R_i$ define $\hat{r}_{i,\lambda}$ be the $0/1$ random variable denoting whether $R_i$ has been extended in the direction $\lambda$. For every point $p$ define random variable, 
$$D_p = \displaystyle\sum_{(i,\lambda)\in S_p}\hat{r}_{i,\lambda}+d_p.$$

Now, we would like to show probabilistically for every point $p$, $D_p$ is at most $d$ with high probability. More formally, we will apply Chernoff bounds to show that $D_p$ does not deviate too much from its expected value. The expected value of $D_p$ is at most $(1-\epsilon)(d-d_p)+d_p$ from the constraints of the linear program. Also, $d_p \le d$, thus we assume $\mu_p = (1-\epsilon)d+\epsilon d_p$.

We would like to lower bound $\Pr[\cap_p\, (D_p \le (1+\delta_p)\mu_p)]$ for some value of $\delta_p$ that we are going to state shortly. Instead, we will bound the probability of the complementary event, i.e., $\Pr[\cup_p\, (D_p > (1+\delta_p)\mu_p)]$. This can be upper bounded by the union bound.
$$\Pr\left[\displaystyle\bigcup_p\; \left(D_p > (1+\delta_p)\mu_p\right)\right] \le \displaystyle\sum_{p}\Pr\left[D_p > (1+\delta_p)\mu_p\right]$$


We apply Chernoff bound, taking $\delta_p=d/\mu_p -1$ and $\mu_p$ as defined earlier. 
Note that the values of $\mu_p$ and $\delta_p$ differs for every point $p$ in $G$.
$$\Pr[D_p > (1+\delta_p)\mu_p] \le exp(\frac{-\mu_p\delta_p^2}{3})$$

We would like the RHS to be at most $1/n^3$. As we assume, from the theorem statement, that $d_p\le (1-\alpha)d$ for a fixed positive fraction $\alpha$ which is independent of $d$, we express $d_p = (1-\alpha_p)d$, where $\alpha_p\ge \alpha$ is an appropriate positive fraction independent of $d$, and the following inequality holds
$$d \ge 9\ln n \frac{(1-\epsilon\alpha_p)}{\epsilon^2\alpha_p^2}$$ 
then $\Pr[D_p > d]\le 1/n^3$. Therefore, the probability of any point having density more than $d$ is at most $1/n$, summing over all the $O(n^2)$ grid points and applying the union bound provided 
$$d \ge 9\ln n \frac{(1-\epsilon\alpha)}{\epsilon^2\alpha^2}$$ 
where, $\alpha=\min_p \alpha_p$.

Finally, we prove that the solution returned is near optimal with high probability. Here again the expected solution size is $(1-\epsilon)|\OPT_{LP}|$. We would like to upper bound $\Pr[|\mathcal{A}|<(1-\epsilon)^2 |\OPT|]$. Applying Chernoff bound, again this is $exp(-\epsilon^2(1-\epsilon)|\OPT_{LP}|/3)$, and we would like this to be at most $1/n$. Thus, $|\OPT_{LP}| \ge \frac{3}{\epsilon^2(1-\epsilon)}\ln n$ and the RHS tends to 0 with $\epsilon$. Hence, $|\mathcal{A}| \ge (1-\epsilon)^2|\OPT_{LP}|$ with probability at least $1-1/n$ provided $|\OPT_{LP}| \ge  \Omega(\frac{\ln n}{\epsilon^2})$. As the density is already assumed to be $\Omega(\ln n\frac{(1-\epsilon\alpha)}{\epsilon^2\alpha^2})$ and the optimum solution is at least the density, the assumption over $\OPT_{LP}$ is redundant. This completes the proof.
\qed

\section{Parameterized Algorithms and Hardness}

In this section, we consider the \textsc{$d$-Runaway Rectangle Escape} problem parameterized by $k$. Note that if the input density is greater than $d$, then we have a trivial \NO{}-instance. On the other hand, we show that if the input density is at most $(d-1)$, then the problem is \FPT{}. Unfortunately, this algorithm does not immediately extend to accommodate the situation when the input may have points of density~    $d$.

We also consider a natural variation of this problem, for which we obtain a parameterized hardness result. Let us call a rectangle \emph{internal} if none of its sides coincide with the boundaries of $R$. We introduce the following question:

\begin{framed}
\noindent \name{{\sc $d$-Constrained Runaway Rectangle Escape}}\\
\noindent Input: A set of $n$ rectangles $\SSS$ in a rectangular region $R$, and integers $p,q$.\\
Question: Is it possible to extend at least $p$ internal rectangles along the horizontal axis, and at least $q$ internal rectangles along the vertical axis, such that the extended configuration has density at most $d$?\\
Parameter: $p+q$
\end{framed}

We show that this particular variant is in fact \WOH{} by a reduction from \textsc{Multi-Colored Clique}, even when $d = 2$. 

\paragraph{Fixed-Parameter Tractability.} Let $(R,\SSS,k,d)$ be an instance of \textsc{$d$-Runaway Rectangle Escape}, where the input density is at most $(d-1)$. We recall that this problem is \NPC{} since the problem was shown, in~\cite{AEYZ13} to be \NPC{} for $d=2, k=n$ even when all the rectangles are disjoint.

As with the approximation algorithm in the previous section, we consider the projections of the input rectangles on the $x$-axis and arrange them according to their left endpoints. Choose a maximum independent set among these intervals by greedily choosing the intervals that end the earliest (and eliminating intervals that overlap with the chosen one). Let $\{X_1, \ldots, X_p\}$ be such an independent set, arranged according to their right endpoints, and let $\mathcal{T} := \{T_1, \ldots, T_p\}$ denote the rectangles from $\SSS$ corresponding to this independent set. Let $a_i$ and $b_i$ denote, respectively, the left and right endpoints of $X_i$. Note that if $p \geq k$, then we may return \YES{} at this point, since the rectangles in $\chi$ can be extended upwards without any mutual conflicts, and this extension will have density at most $d$ because input density is at most $(d-1)$ to begin with. Therefore, $p < k$.

We repeat this process on the $y$-projections of the rectangles, and let $\mathcal{T}^\prime := \{T_1^\prime, \ldots, T_q^\prime\}$ denote the rectangles from $\SSS$ corresponding to the independent set obtained in this case. We let $(a_i^\prime,b_i^\prime)$ denote the top and bottom endpoints of the $y$-projection of $T_i^\prime$. Again, we may assume that $q < k$, otherwise we are done. Now consider the lines given by $x = b_i$ for $1 \leq i \leq p$ and $y = b_j^\prime$ for $1 \leq j \leq q$. Let $g(i,j)$ denote the intersection of the lines $x = b_i$ and $y = b_j^\prime$. For every rectangle $H$ in $\SSS$, observe that there exists $1 \leq i \leq p$ and $1 \leq j \leq q$ such that $H$ contains $g(i,j)$. Indeed, suppose not. Then this would imply, for instance, that $H$ is not stabbed by any of the vertical lines $x = b_i$, which implies that there exists $i \in [p]$ for which the left endpoint of the $x$-projection of $H$ is after $b_i$, and the right endpoint is before $b_{i+1}$. However, this contradicts the greedy construction of $\chi$. A similar argument can be made for the horizontal lines. 

Now, we have a collection of less than $k^2$ points that pierce all the rectangles in $\SSS$. Since the input density was at most $(d-1)$, each $g(i,j)$ can be contained in at most $(d-1)$ of the input rectangles. Therefore, the total number of rectangles is at most $(dk^2)$. We may now guess the subset of $k$ rectangles that we would like to extend in time: $${dk^2 \choose k} \leq \left(\frac{dk^2e}{k}\right)^k = (dke)^k.$$

For each guess, we can further guess the direction of the extension of the chosen rectangles --- noting that there are at most four possibilities for each rectangle, this is an additional overhead of $4^k$. Note that we spend polynomial time in identifying the stabbing lines (and resolving the instance at that stage if it is called for). So overall, the running time of our algorithm is $O((dke)^k4^k)n^{O(1)}$. 

\begin{theorem}
\label{thm:fpt}
The \textsc{$d$-Runaway Rectangle Escape Problem} can be resolved in time $2^{O(k \log k)}n^{O(1)}$ when the input configuration has density at most $(d-1)$.
\end{theorem}

Note that the difficulty with input configurations that have points of density $d$ is that having an independent set of size $k$ on the $x$-projections does not imply that we have a solution, because quite possibly many of the rectangles in the independent set are ``blocked'' by points of density $d$. Even if we forbid such rectangles upfront, as in the previous section, we may have an unbounded number of rectangles that are stuck horizontally or vertically. As it turns out, the rectangles that are stuck horizontally \emph{and} vertically pose no problems, because they can be declared forbidden and eliminated from the search space. Similarly, the number of rectangles that are not stuck in either direction can be bounded by $(k^2d)$ by an argument along the lines of what we had for Theorem~\ref{thm:fpt}. However, there may be an unbounded number of rectangles that are stuck only vertically (or only horizontally), and this is where the argument for Theorem~\ref{thm:fpt} does not extend to the case when the input density is $d$.

\paragraph{W-hardness} We now turn to the \name{$2$-Constrained Runaway Rectangle Escape}. Let $(G,V,k)$ be an instance of \name{Multi-Colored Clique}, where the partitions of $V$ are given by $V_1 \uplus \cdots \uplus V_k$. We assume, without loss of generality, that all the parts have the same number of vertices. We denote the vertices in $V_i$ by $v_i[1], \ldots, v_i[t]$. 

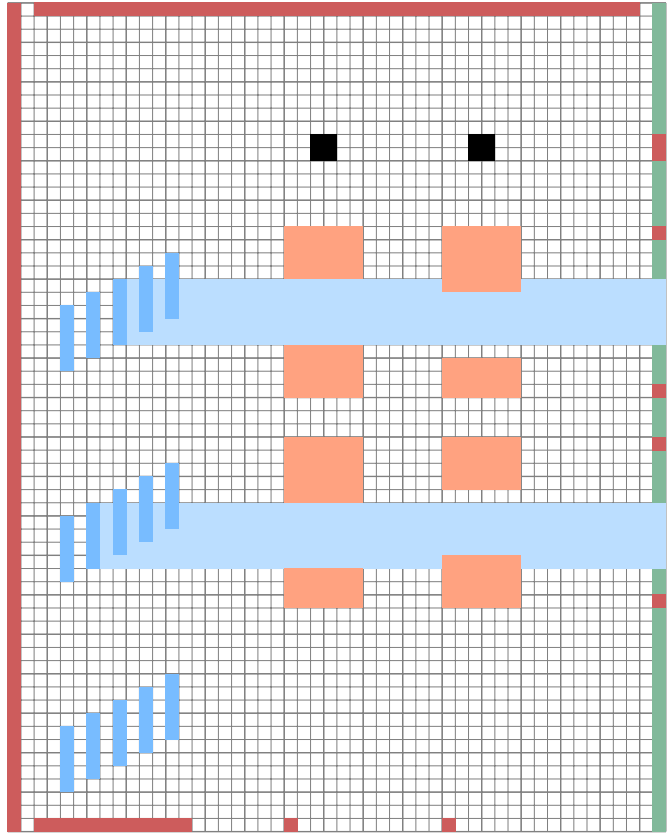
\begin{figure}[H]

\begin{center}
\begin{tikzpicture}[scale=0.35]

\draw (0,-1.5) to[grid with coordinates] (25,30);

\draw [fill=SeaGreen!60,draw=none] (24.5,-1.5) rectangle (25,30);
\begin{scope}[yshift=-1.5cm,xshift=0.5cm]

\draw [fill=DodgerBlue!60,draw=none] (1.5,1.5) rectangle (2,4);
\draw [fill=DodgerBlue!60,draw=none,xshift=1cm,yshift=0.5cm] (1.5,1.5) rectangle (2,4);
\draw [fill=DodgerBlue!60,draw=none,xshift=2cm,yshift=1cm] (1.5,1.5) rectangle (2,4);
\draw [fill=DodgerBlue!60,draw=none,xshift=3cm,yshift=1.5cm] (1.5,1.5) rectangle (2,4);
\draw [fill=DodgerBlue!60,draw=none,xshift=4cm,yshift=2cm] (1.5,1.5) rectangle (2,4);
\end{scope}

\begin{scope}[yshift=6.5cm,xshift=0.5cm]

\draw [fill=DodgerBlue!30,draw=none,xshift=1cm,yshift=.5cm] (1.5,1.5) rectangle (23.5,4);

\draw [fill=OrangeRed!50,draw=none,xshift=6cm,yshift=1cm] (10,4) rectangle (13,6);
\draw [fill=OrangeRed!50,draw=none,xshift=6cm,yshift=-3.5cm] (10,4) rectangle (13,6);

\draw [fill=OrangeRed!50,draw=none,yshift=1cm] (10,3.5) rectangle (13,6);
\draw [fill=OrangeRed!50,draw=none,yshift=-3.5cm] (10,4) rectangle (13,5.5);

\draw [fill=DodgerBlue!60,draw=none] (1.5,1.5) rectangle (2,4);
\draw [fill=DodgerBlue!60,draw=none,xshift=1cm,yshift=0.5cm] (1.5,1.5) rectangle (2,4);
\draw [fill=DodgerBlue!60,draw=none,xshift=2cm,yshift=1cm] (1.5,1.5) rectangle (2,4);
\draw [fill=DodgerBlue!60,draw=none,xshift=3cm,yshift=1.5cm] (1.5,1.5) rectangle (2,4);
\draw [fill=DodgerBlue!60,draw=none,xshift=4cm,yshift=2cm] (1.5,1.5) rectangle (2,4);
\end{scope}

\begin{scope}[yshift=14.5cm,xshift=0.5cm]
\draw [fill=DodgerBlue!30,draw=none,xshift=2cm,yshift=1cm] (1.5,1.5) rectangle (22.5,4);

\draw [fill=OrangeRed!50,draw=none,yshift=1cm] (10,4) rectangle (13,6);
\draw [fill=OrangeRed!50,draw=none,yshift=-3.5cm] (10,4) rectangle (13,6);

\draw [fill=OrangeRed!50,draw=none,xshift=6cm,yshift=1cm] (10,3.5) rectangle (13,6);
\draw [fill=OrangeRed!50,draw=none,xshift=6cm,yshift=-3.5cm] (10,4) rectangle (13,5.5);

\draw [fill=DodgerBlue!60,draw=none] (1.5,1.5) rectangle (2,4);
\draw [fill=DodgerBlue!60,draw=none,xshift=1cm,yshift=0.5cm] (1.5,1.5) rectangle (2,4);
\draw [fill=DodgerBlue!60,draw=none,xshift=2cm,yshift=1cm] (1.5,1.5) rectangle (2,4);
\draw [fill=DodgerBlue!60,draw=none,xshift=3cm,yshift=1.5cm] (1.5,1.5) rectangle (2,4);
\draw [fill=DodgerBlue!60,draw=none,xshift=4cm,yshift=2cm] (1.5,1.5) rectangle (2,4);

\end{scope}

\draw [fill=IndianRed,draw=none] (0,-1.5) rectangle (.5,30);
\draw [fill=IndianRed,draw=none] (1,-1.5) rectangle (7,-1);
\draw [fill=IndianRed,draw=none] (1,29.5) rectangle (24,30);

\draw [fill=IndianRed,draw=none,yshift=15cm] (24.5,0) rectangle (25,.5);
\draw [fill=IndianRed,draw=none,yshift=21cm] (24.5,0) rectangle (25,.5);

\draw [fill=IndianRed,draw=none,yshift=7cm] (24.5,0) rectangle (25,.5);
\draw [fill=IndianRed,draw=none,yshift=13cm] (24.5,0) rectangle (25,.5);

\draw [fill=IndianRed,draw=none,yshift=24cm] (24.5,0) rectangle (25,1);

\draw [fill=IndianRed,draw=none,xshift=-8cm,yshift=-1.5cm] (24.5,0) rectangle (25,.5);
\draw [fill=IndianRed,draw=none,xshift=-14cm,yshift=-1.5cm] (24.5,0) rectangle (25,.5);

\draw [fill=Black,draw=none,xshift=-12.5cm,yshift=24cm] (24,0) rectangle (25,1);

\draw [fill=Black,draw=none,xshift=-6.5cm,yshift=24cm] (24,0) rectangle (25,1);

\end{tikzpicture}
\end{center}
\caption{A cross-section schematic, \emph{not} drawn to scale, of the reduction from Multi-Colored Clique. The groups of blue rectangles correspond to vertices from a particular partition in the instance of Multi-Colored Clique. The red rectangles indicate two overlapping rectangles placed along the borders of $R$, while the green rectangle is a single rectangle, again aligned to the right border of $R$. The orange rectangles are the incidence gadgets and the black rectangles correspond to the edges.}
\label{fig:vertexselection}
\end{figure}

The reader is referred to Figure~\ref{fig:vertexselection} for a schematic of the construction that we are going to describe. We first introduce the rectangles corresponding to vertices, which we call \emph{selection gadgets}. Every vertex in $V$ is associated with a rectangle of unit width and height $(t+1)$. We use $T_i[j]$ to refer to the rectangle corresponding to the vertex $v_i[j]$, where $1 \leq i \leq k$ and $1 \leq j \leq t$. 

We place the bottom-left corner of $T_i[j]$ at $(2+2j,3+j+(2t+5)(i-1))$. This is simply a collection of $t$ rectangles cascading successively in the top-right direction, with the collection of rectangles for vertices in $V_i$ appropriately offset from the collection corresponding to vertices in $V_{i+1}$.  We use $\mathcal{T}_i$ to refer to $\{T_i[j] ~|~ j \in [t]\}$. Note that at most two of the rectangles from any $\mathcal{T}_i$ can be extended either to the right or the left (since $d = 2$ and all of these rectangles are stabbed by a single horizontal line). We will refine this observation further with the help of additional rectangles, to ensure that at most one of them can be extended to the right, and none of them in any of the other directions. 

We now turn to the edges in $G$. For each $e_\ell \in G$, we introduce an unit square $T_\ell$ with its lower left corner at $(3t + 12\ell,(2t+5)^2)$. Informally, all the squares corresponding to the edges are placed on one horizontal line, suitably spaced out. Further, the $y$-coordinates of their lower-left corners are large enough to ensure that the squares are placed above all the vertex gadgets.

Next, we add incidence gadgets. These rectangles ensure that if $T_i[j]$ is extended to the right and $T_\ell$ is extended downwards, then the edge $e_\ell$ is incident to $v_i[j]$. We first informally describe the setup. Let $B_a[r]$ denote the rectangle obtained by extending $T_a[r]$ towards the right.  Let $e_\ell = (v_a[p],v_b[q])$. We will place two rectangles $W_a[p],Z_a[p]$ to the right of $\mathcal{T}_a$ and below $T_\ell$. The rectangle $W_a[p]$ will intercept the bands $B_a[r]$ for $r > p$, but will not overlap $B_a[p]$, and the rectangle $Z_a[p]$ will intercept the bands $B_a[r]$ for $r < p$, but again will not overlap $B_a[p]$. This ensures that if $T_a[r]$ is extended to the right and $e_\ell$ is not incident to $r$, then a point of density two is created by the overlap of either $W_a[p]$ or $Z_a[p]$ with the extended rectangle $T_a[r]$, thus forbidding $T_\ell$ from being extended downwards. This process is repeated for the collection $\mathcal{T}_b$. 

Formally, for every edge $e_\ell = (v_a[p],v_b[q])$, we place the following four rectangles, which we call \emph{incidence gadgets}. All these rectangles are seven units wide, and the $x$-coordinate of their lower-left corner is three units less than the $x$-coordinate of the lower-left corner of $T_\ell$. That is, if we consider the $x$-projections of these four rectangles along with the $x$-projection of $T_\ell$, then we will find the $x$-projection of $T_\ell$ exactly at the center, and the remaining four intervals coinciding. We now describe how $W_a[p], Z_a[q]$ are placed along the $y$-axis, and note that the rectangles $W_b[q]$ and $Z_b[q]$ are placed similarly. 

\begin{enumerate}
\item The rectangle $W_a[p]$. The $y$-coordinate of the bottom edge of $W_a[p]$ is the same as the $y$-coordinate of the upper edge of $T_a[p]$. The $y$-coordinate of the top edge of $W_a[p]$ is two more than the $y$-coordinate of the upper edge of $T_a[t]$, that is, we make sure that this rectangle ``juts out'' over and above the last rectangle in the group $\mathcal{T}_a$. This will be useful later, when we would like to forbid this rectangle from extending to the right, \emph{without} forbidding any of the rectangles in $\mathcal{T}_i$ from extending to the right. 
\item The rectangle $Z_a[p]$. The $y$-coordinate of the top edge of $Z_a[p]$ is the same as the $y$-coordinate of the bottom edge of $T_a[p]$. The $y$-coordinate of the bottom edge of $Z_a[p]$ is two less than the $y$-coordinate of the bottom edge of $T_a[1]$.
\end{enumerate}

We now incorporate some rectangles along the boundary, which we will refer to as \emph{guards}. The purpose here is to ``block'' certain extensions. To begin with, we place two overlapping unit-width rectangles along the entire left boundary, and two overlapping unit-height rectangles along the top boundary of $R$. This ensures, for example, that none of the internal rectangles can be extended to either the left or the top. Further, we place a single rectangle, denoted by $H$, that covers the entire right boundary (stopping short of the guard rectangles on top to avoid a region of density three).

Next, we would like to ensure that the rectangles $T_i[j]$ can only be extended to the right. To this end, we place two overlapping rectangles of unit height along the bottom boundary of $R$, wide enough to block any $T_i[j]$ from extending downwards, for $1 \leq i \leq k$ and $1 \leq j \leq t$. Specifically:

\begin{itemize}
\item The $x$-coordinate of the bottom left corner of these rectangles is one less than the $x$-coordinate of the bottom left corner of $T_1[1]$.
\item The $x$-coordinate of the bottom right corner of these rectangles is one more than the $x$-coordinate of the bottom right corner of $T_1[t]$.
\end{itemize}

We add a unit square on the right boundary (overlapping $H$), whose lower-right corner has the $y$-coordinate $(2t+5)^2$. This effectively blocks the rectangles corresponding to the edges from extending to the right. 

Finally, we add rectangles along the bottom and right boundaries to ensure that the rectangles in the incidence gadgets are blocked from being extended to either the right or the downwards. In this context, we introduce unit squares $H_1,\ldots, H_k$ and $H_1^\dagger,\ldots, H_k^\dagger$, to be placed along the right boundary. The $y$-coordinate of the upper-right corner of the square $H_i$ is two more than the $y$-coordinate of the upper edge of $T_i[t]$. This, together with $H$, ensures that the rectangles $W_i[j]$ are blocked from extending towards the right, for any $1 \leq i \leq k$ and $1 \leq j \leq t$. Similarly, the $y$-coordinate of the lower-right corner of the square $H_i^\dagger$ is two less than the $y$-coordinate of the bottom edge of $T_i[1]$. Again, together with $H$, this ensures that the rectangles $Z_i[j]$ are blocked from extending towards the right, for any $1 \leq i \leq k$ and $1 \leq j \leq t$. Note that these rectangles do not block any rectangles in ${\mathcal T}_i$ from extending to the right, because of their unit height. 

For this instance, we let $p = k$ and $q = {k \choose 2}$. This completes the description of the construction, and we now turn to a proof of correctness. It is useful to keep in mind that the guards are the only rectangles that are not internal. 

In the forward direction, let $c_1, \ldots, c_t$, $c_i \in [t]$, be such that the vertices $v_i[c_i]$ form a multi-colored clique. We then extend the rectangles $T_i[c_i]$ to the right, and the unit squares $T_\ell$ corresponding to the edges of the clique downwards. It is easy to check that the guards do not interfere with any of these extensions, that is, there are no points of density three on the boundary after these rectangles are extended as described. Also, extending the rectangles from the selection gadgets alone creates no points of density greater than two. We now address the edge extensions. Let $e_\ell = (v_i[c_i],v_j[c_j])$ be an edge in the clique. Observe that the rectangles in the incidence gadget corresponding to the rectangle $T_{\ell}$ skirt the edges of the bands $B_i[c_i]$ and $B_j[c_j]$, without overlapping them. Therefore, it can be verified that we create no points of density greater than two when the square $T_\ell$ is extended downwards. 

In the reverse direction, we observe that at most one rectangle can be extended to the right from ${\mathcal T_i}$, and none of them can be extended to the left. Further, none of the other internal rectangles can be extended along the horizontal axis while maintaining density at most two. Since we have to extend at least $k$ rectangles along the horizontal axis, it follows that any solution extends exactly one rectangle from each ${\mathcal T_i}$, for $1 \leq i \leq k$. Let $1 \leq c_i \leq t$ be such that $T_i[c_i]$ was the rectangle that was extended to the right. We claim that the vertices $v_i[c_i]$ form a multi-colored clique in $G$. Indeed, observe that if $T_\ell$ is extended downwards, where $e_\ell = (v_i[p],v_j[q])$, then the rectangle extended from $\mathcal{T}_i$ must be $T_i[p]$ and the rectangle extended from $\mathcal{T}_j$ must be $T_j[q]$ --- indeed, the extension of any other rectangle from either collection will lead to a point of density three (combined with the incidence gadgets for $T_\ell$. Therefore, a rectangle corresponding to an edge can be extended downwards only if it is an edge from $G[\{v_1[c_1], \ldots, v_t[c_t]\}]$. Recall that the guard vertices are positioned so that none of the internal rectangles can be extended upwards, and only the squares corresponding to the edges can be extended downwards. Therefore, if the claimed subgraph does not induce a clique, we conclude that the solution falls short of the ${k \choose 2}$ extensions that were required along the vertical axis. Thus, we have shown the following.

\begin{theorem}
The \name{$2$-Constrained Runaway Rectangle Escape} is $W[1]$-hard. 
\end{theorem}

\section{The Square Escape Problem}

In this section, we look in to a special case of the \RE{}, where the rectangular region $R$ is given as a grid of unit squares, and every rectangle is a unit square aligned to the grid. This is same as having grid points instead of squares and orthogonal line segments joining the grid points to the boundary of $R$ instead of extensions of the squares. As it turns out this problem, in the latter guise, has been studied for unit density and an $O(n^2\log n)$ time algorithm is devised in~\cite{Palios:1997:CMN:249088.249092}. We show that despite being a rather severely specialized version of \RE{}, even this formulation is \NPH{} for density two. In particular, the problem of determining if all the squares can be extended while maintaining density two is \NPH{}, while the ``runaway'' version enjoys an improved approximation algorithm and is fixed-parameter tractable irrespective of the density of the input squares.

\begin{framed}
\noindent \name{{\sc $d$-Runaway Square Escape}}\\
\noindent Input: A set $\SSS$ of $n$ squares from an $m \times m$ grid $R$, and an integer $k$.\\
Question: Is $\rho(\SSS,d) \geq k$?
\end{framed}

We first show that the \name{$d$-Runaway Square Escape} is \NPH{} even when $k = n$ and $d = 2$. We reduce from the version of \name{Not-All-Equals SAT}. It is known that not-all-equals satisfiability continues to be \NPC{} for this restricted formulation~\cite{AD09}. 

\begin{framed}
\noindent \name{\sc Not-All-Equals SAT}\\
\noindent Input: A conjunction of clauses each containing 2 or 3 variables, all in their positive form, and every variable appears in at most 3 clauses.\\
Question: Does their exists a satisfying assignment $\phi$ such that every clause has at least one variable to be true and at least another variable to be false?
\end{framed}

Before we describe the construction, we introduce some terminology. For a square $s$ located on the $i^{th}$ row and the $j^{th}$ column of the given grid, we use $R(s)$ and $C(s)$ to refer to $i$ and $j$, respectively. We say we place a square at $(i,j)$ to indicate that a square is placed in the location determined by the intersection of the $i^{th}$ row and $j^{th}$ column. 

Suppose we are working on an $m \times m$ grid. When we say that we \emph{block} a square $s$, say, in the upward direction, then this means that we introduce two overlapping squares at $(m,C(s))$, if they are not already present. We are only allowed to block a square $s$ if there are either no squares at $(m,C(s))$, or if there are two squares at $(m,C(s))$. The terminology is motivated by the fact that when we block a square $s$ in the upward direction, no extension of density at most two can extend $s$ in the upward direction. 

When we say that we \emph{partially block} a square $s$ in the upward direction, then this means that we introduce one square at $(m,C(s))$, if not already present. In particular, a square on column $j$ cannot be partially blocked if there are two squares placed already at $(m,j)$.  These squares are called \emph{guards}.

We let $\phi$ denote an instance of \name{Not-All-Equals SAT} where every clause has two or three variables and every variable occurs in at most three clauses. Let $v_1,\ldots,v_n$ be the variables involved in $\phi$, and let $C_1, \ldots, C_m$ denote the clauses of $\phi$. 

For every variable, we will introduce three squares corresponding to the variable, which we simply call the variable gadget. We then add more squares to ensure that these three squares are always extended in the same direction, and these collections of squares are called the copy gadgets. Finally, we add three squares for every clause, which we call the clause gadgets. 

For a variable $v_i$, let $\mathcal{V}_i := \{s_i[1],s_i[2],s_i[3]\}$ denote the three squares involved in the corresponding variable gadget. We say we place $\mathcal{V}_i$ at $(x,y)$ to mean that $s_i[1]$ is placed at $(x,y+4)$, $s_i[2]$ is placed at $(x+2,y+2)$ and $s_i[3]$ is placed at $(x+4,y)$. The envelope of a variable gadget that is placed at $(x,y)$, denoted by $\mathcal{E}_i$, is defined as the rectangular region whose lower-left corner is at $(x,y)$ and whose upper-right corner is at $(x+25,y+25)$. All the squares that participate in the copy gadget for $\mathcal{V}_i$ will be placed in $\mathcal{E}_i$. We now describe the individual components of the construction.

\paragraph{Variable Gadgets} The variable gadget corresponding to $v_1$ is placed at $(0,0)$. The variable gadget corresponding to $v_i$ is placed at the top-right corner of $\mathcal{E}_{i-1}$, for $2\le i \le n$. All the squares in the variable gadgets are blocked downwards and to their left, while they are partially blocked upwards and to their right. 

\paragraph{Clause Gadgets} Let $C_1, \ldots, C_m$ be an arbitrary but fixed ordering of the clauses. Let $C_j = \{v_{i_1},v_{i_2},v_{i_3}\}$ be a clause of length three. Within a clause, we order the variables according to increasing order of their indices. Let $C_j$ be the $f_j[x]^{th}$ clause that $v_{i_x}$ appears in. For example, for a clause $C_3 := \{v_2,v_3,v_7\}$, we may have $f_3[1] = 2$ to denote the fact that $C_3$ is the second clause that $v_2$ appears in. Note that $f_j[x] \in \{1,2,3\}$ for the particular instance of \name{Not-All-Equals SAT} that we have started with. 

For this clause $C_j$, we introduce squares $t_j[1]^U, t_j[2]^U,t_j[3]^U$ and $t_j[1]^R, t_j[2]^R,t_j[3]^R$, placed in the following manner.

\begin{enumerate}
\item For $x \in \{1,2,3\}$, the square $t_j[x]^U$ is placed in the same column as $s_{i_x}[f_j[x]]$. The row it is placed in is $2j + 25n + 10$. In particular, it is $2j+10$ units above $\mathcal{E}_n$. 
\item For $x \in \{1,2,3\}$, the square $t_j[x]^R$ is placed in the same row as $s_{i_x}[f_j[x]]$. The column it is placed in is $2j + 25n + 10$. In particular, it is $2j+10$ units to the right of $\mathcal{E}_n$. 
\end{enumerate}

If we have a clause of length two, then we place squares $t_j[1]^U, t_j[2]^U$ and $t_j[1]^R, t_j[2]^R$ exactly as described above. Further, we add two dummy squares $P$ and $Q$, where $P$ is placed on the same row as $t_j[1]^U, t_j[2]^U$, and is placed on an empty column $c$ such that $C(t_j[1]^U) < c < C(t_j[2]^U)$. Similarly, $Q$ is placed on the same column as $t_j[1]^R$, and is placed on an empty row $r$ such that  $R(t_j[1]^U) < r < R(t_j[2]^U)$. The square $P$ is blocked up and down, while the square $Q$ is blocked on the right and left. We note that if empty rows or columns are not available, then the spacing between the envelopes of the variables can be easily adjusted to free up space. We do not incorporate this detail in the interest of a simpler presentation. 

\begin{figure}[h]

\begin{center}
\begin{tikzpicture}[scale=0.25]

\draw [fill=DodgerBlue!50,draw=none,xshift=3cm,yshift=3cm] (1,5) rectangle (27,6);
\draw [fill=DodgerBlue!50,draw=none,xshift=3cm,yshift=3cm] (3,3) rectangle (27,4);
\draw [fill=DodgerBlue!50,draw=none,xshift=3cm,yshift=3cm] (5,1) rectangle (27,2);

\draw [fill=Black,draw=none,xshift=-1cm,yshift=3cm] (1,5) rectangle (2,6);
\draw [fill=Black,draw=none,xshift=-1cm,yshift=1cm] (1,5) rectangle (2,6);
\draw [fill=Black,draw=none,xshift=-1cm,yshift=-1cm] (1,5) rectangle (2,6);

\draw [fill=Black,draw=none,xshift=7cm,yshift=-5cm] (1,5) rectangle (2,6);
\draw [fill=Black,draw=none,xshift=5cm,yshift=-5cm] (1,5) rectangle (2,6);
\draw [fill=Black,draw=none,xshift=3cm,yshift=-5cm] (1,5) rectangle (2,6);

\draw [fill=DodgerBlue,draw=none,xshift=3cm,yshift=3cm] (1,5) rectangle (2,6);
\draw [fill=DodgerBlue,draw=none,xshift=3cm,yshift=3cm] (3,3) rectangle (4,4);
\draw [fill=DodgerBlue,draw=none,xshift=3cm,yshift=3cm] (5,1) rectangle (6,2);

\draw [fill=SlateGray,draw=none,xshift=28cm,yshift=3cm] (1,5) rectangle (2,6);
\draw [fill=SlateGray,draw=none,xshift=28cm,yshift=1cm] (1,5) rectangle (2,6);
\draw [fill=SlateGray,draw=none,xshift=28cm,yshift=-1cm] (1,5) rectangle (2,6);

\draw [fill=SeaGreen,draw=none,xshift=20cm,yshift=3cm] (1,5) rectangle (2,6);
\draw [fill=OrangeRed,draw=none,xshift=20cm,yshift=1cm] (1,5) rectangle (2,6);
\draw [fill=SeaGreen,draw=none,xshift=20cm,yshift=-1cm] (1,5) rectangle (2,6);

\end{tikzpicture}
\end{center}

\caption{A schematic of the clause gadget.}
\end{figure}
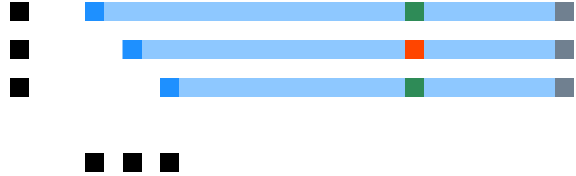

\paragraph{Copy Gadgets} Let $\mathcal{V}_i$ be a variable gadget placed at $(x,y)$. Then we introduce the following squares in the copy gadget corresponding to $\mathcal{V}_i$:

\begin{itemize}
\item We place squares at $(x,y+8), (x+2,y+12), (x+2,y+16), (x+4,y+20)$. Further, we place squares at $(x+8,y+2),(x+12,y+4),(x+16,y),(x+20,y+2)$.
\item We place squares at $(x+8,y+8),(x+12,y+12),(x+16,y+16),(x+20,y+20)$. We call these the \emph{blockers}.
\item For each blocker at $(p,q)$, we place two additional squares at $(p-2,q)$ and $(p,q-2)$. These we call the \emph{anchors}. The anchors at $(p-2,q)$ are blocked upwards and downwards, while the rest of the anchors are blocked to their left and right. 
\end{itemize}

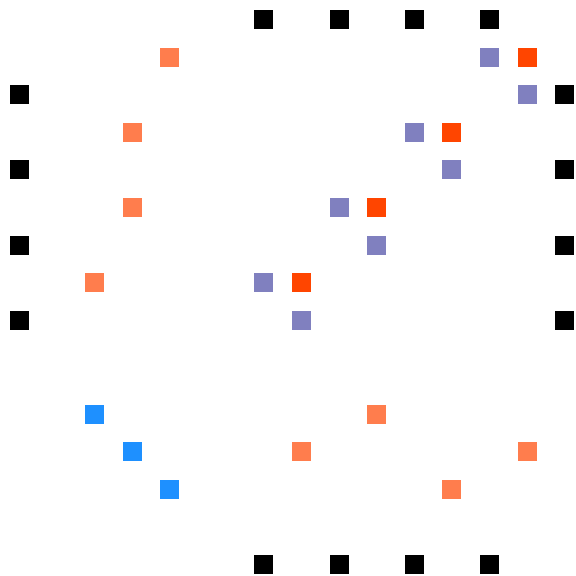
\begin{figure}[H]

\begin{center}
\begin{tikzpicture}[scale=0.25]

\draw [fill=DodgerBlue,draw=none,xshift=3cm,yshift=3cm] (1,5) rectangle (2,6);
\draw [fill=DodgerBlue,draw=none,xshift=3cm,yshift=3cm] (3,3) rectangle (4,4);
\draw [fill=DodgerBlue,draw=none,xshift=3cm,yshift=3cm] (5,1) rectangle (6,2);

\begin{scope}[yshift=3cm]
\draw [fill=OrangeRed!70,draw=none,xshift=3cm,yshift=7cm] (1,5) rectangle (2,6);
\draw [fill=OrangeRed!70,draw=none,xshift=5cm,yshift=11cm] (1,5) rectangle (2,6);
\draw [fill=OrangeRed!70,draw=none,xshift=5cm,yshift=15cm] (1,5) rectangle (2,6);
\draw [fill=OrangeRed!70,draw=none,xshift=7cm,yshift=19cm] (1,5) rectangle (2,6);
\end{scope}

\begin{scope}[xshift=3cm]
\draw [fill=OrangeRed!70,draw=none,xshift=11cm,yshift=1cm] (1,5) rectangle (2,6);
\draw [fill=OrangeRed!70,draw=none,xshift=15cm,yshift=3cm] (1,5) rectangle (2,6);
\draw [fill=OrangeRed!70,draw=none,xshift=19cm,yshift=-1cm] (1,5) rectangle (2,6);
\draw [fill=OrangeRed!70,draw=none,xshift=23cm,yshift=1cm] (1,5) rectangle (2,6);
\end{scope}

\begin{scope}[xshift=3cm,yshift=3cm]
\draw [fill=OrangeRed,draw=none,xshift=11cm,yshift=7cm] (1,5) rectangle (2,6);
\draw [fill=OrangeRed,draw=none,xshift=15cm,yshift=11cm] (1,5) rectangle (2,6);
\draw [fill=OrangeRed,draw=none,xshift=19cm,yshift=15cm] (1,5) rectangle (2,6);
\draw [fill=OrangeRed,draw=none,xshift=23cm,yshift=19cm] (1,5) rectangle (2,6);
\end{scope}

\begin{scope}[xshift=3cm,yshift=3cm]
\draw [fill=NavyBlue!50,draw=none,xshift=9cm,yshift=7cm] (1,5) rectangle (2,6);
\draw [fill=Black,draw=none,xshift=9cm,yshift=21cm] (1,5) rectangle (2,6);
\draw [fill=Black,draw=none,xshift=9cm,yshift=-8cm] (1,5) rectangle (2,6);

\draw [fill=NavyBlue!50,draw=none,xshift=11cm,yshift=5cm] (1,5) rectangle (2,6);
\draw [fill=Black,draw=none,xshift=25cm,yshift=5cm] (1,5) rectangle (2,6);
\draw [fill=Black,draw=none,xshift=-4cm,yshift=5cm] (1,5) rectangle (2,6);

\draw [fill=NavyBlue!50,draw=none,xshift=13cm,yshift=11cm] (1,5) rectangle (2,6);
\draw [fill=Black,draw=none,xshift=13cm,yshift=21cm] (1,5) rectangle (2,6);
\draw [fill=Black,draw=none,xshift=13cm,yshift=-8cm] (1,5) rectangle (2,6);

\draw [fill=NavyBlue!50,draw=none,xshift=15cm,yshift=9cm] (1,5) rectangle (2,6);
\draw [fill=Black,draw=none,xshift=25cm,yshift=9cm] (1,5) rectangle (2,6);
\draw [fill=Black,draw=none,xshift=-4cm,yshift=9cm] (1,5) rectangle (2,6);

\draw [fill=NavyBlue!50,draw=none,xshift=17cm,yshift=15cm] (1,5) rectangle (2,6);
\draw [fill=Black,draw=none,xshift=17cm,yshift=21cm] (1,5) rectangle (2,6);
\draw [fill=Black,draw=none,xshift=17cm,yshift=-8cm] (1,5) rectangle (2,6);

\draw [fill=NavyBlue!50,draw=none,xshift=19cm,yshift=13cm] (1,5) rectangle (2,6);
\draw [fill=Black,draw=none,xshift=25cm,yshift=13cm] (1,5) rectangle (2,6);
\draw [fill=Black,draw=none,xshift=-4cm,yshift=13cm] (1,5) rectangle (2,6);

\draw [fill=NavyBlue!50,draw=none,xshift=21cm,yshift=19cm] (1,5) rectangle (2,6);
\draw [fill=Black,draw=none,xshift=21cm,yshift=21cm] (1,5) rectangle (2,6);
\draw [fill=Black,draw=none,xshift=21cm,yshift=-8cm] (1,5) rectangle (2,6);

\draw [fill=NavyBlue!50,draw=none,xshift=23cm,yshift=17cm] (1,5) rectangle (2,6);
\draw [fill=Black,draw=none,xshift=25cm,yshift=17cm] (1,5) rectangle (2,6);
\draw [fill=Black,draw=none,xshift=-4cm,yshift=17cm] (1,5) rectangle (2,6);

\end{scope}

\end{tikzpicture}
\end{center}
\caption{A general schematic of a copy gadget.}
\label{square}
\end{figure}

The variable gadgets, their corresponding copy gadgets, and the clause gadgets, together comprise the reduced instance. We now argue the equivalence of the two instances. 

In the forward direction, let $\tau: \{v_1,\ldots,v_n\} \rightarrow \{0,1\}$ be a not-all-equals satisfying assignment. If $\tau(v_i) = 1$, then we extend all the squares in $\mathcal{V}_i$ to the right, and if $\tau(v_i) = 0$, then we extend all the squares in $\mathcal{V}_i$ upwards. It can be shown that all the squares in the copy gadgets continue to have a valid extension (see Figure~\ref{square} onwards for illustration). It is important to ensure here that for any fixed column (or row), we extend at most one square upwards (or rightwards) along that column (or row). This ensures that all the ``crossings'' encountered when we proceed to extend the squares corresponding to clause gadgets have density at most two. Also, extending two squares to the left or downwards along any row or column causes no problems, because any potential interference comes from clause gadgets being extended (respectively) downwards or to the left --- but the placements of these gadgets are such that these extensions are guaranteed to be parallel, and consequently, non-crossing. 

Among the squares $t_j[1]^U, t_j[2]^U,t_j[3]^U$, notice that at most two of them are in locations with density two because at most two of the corresponding squares in the variable gadgets were extended upwards (recall that we start with a not-all-equals satisfying assignment). Therefore, we extend the square that is free upwards, and the other two to the left and right, respectively. A similar argument works for the squares $t_j[1]^R, t_j[2]^R,t_j[3]^R$. Finally, all the guards can be trivially extended to the edge that they are the closest to. 

In the reverse direction, we first note that in any valid extension, all the squares in a variable gadget must be extended in the same direction. For example, for any $1 \leq i \leq n$, if $s_i[1]$ and $s_i[2]$ are extended upwards and to the right respectively, then there are corresponding anchor squares that are forced to be extended to the right and upwards, which then create a point of density three at the corresponding blocker square. It can be argued, therefore, that $s_i[1]$ and $s_i[2]$ must be extended in the same direction, and similarly, that $s_i[2]$ and $s_i[3]$ must be extended in the same direction. It follows that all three of them must be extended in the same direction --- and since they are blocked on the left and downwards, they must be extended either upwards or to the right.

We suggest an assignment to the variables of $\phi$ as follows. If the squares in $\mathcal{V}_i$ are extended to the right, then we set $v_i$ to $1$ and to $0$ otherwise. If this is not a valid not-all-equals assignment, then consider the squares corresponding to a violated clause. Assume, without loss of generality, that all variables in this clause were set to one, therefore, the squares corresponding to the variables were extended to the right. If this was a clause of length three, then observe that the square in the clause gadget corresponding to the second variable is now blocked in all four directions (recall that the squares corresponding to the variables were partially blocked on the right), and cannot be extended. If this was a clause of length two, then the dummy square $Q$ corresponding to the clause is similarly blocked in all four directions. In all four cases, we get the desired contradiction. Thus we have shown the following.
\begin{theorem}
\name{$d$-Runaway Square Escape} is \NPC{} even when $d = 2$ and $k = n$.
\end{theorem}

On the other hand, we know that for the {$d$-Runaway Square Escape Problem}, we may find the maximum number of rectangles that can be extended vertically in polynomial time. Indeed, for every column, we extend the $d$ squares ``closest to the top'' upwards, and the $d$ squares ``closest to the bottom'' downwards. This is evidently an optimal solution. A similar argument holds for finding the maximum number of rectangles that can be extended horizontally. Since any solution that extends the squares in any of the four directions extends at least half of the squares either vertically or horizontally, we have a simple two-approximation algorithm. It is also easy to check that the fixed-parameter tractable algorithm described for rectangles works for squares with no assumptions on the density of the input configuration.

\begin{figure}[H]

\begin{center}
\begin{tikzpicture}[scale=0.25]
\label{copy1}

\begin{scope}[xshift=3cm]
\draw [fill=OrangeRed!70,draw=none,xshift=11cm,yshift=1cm] (0.7,-1) rectangle (2.3,6);
\draw [fill=OrangeRed!70,draw=none,xshift=15cm,yshift=3cm] (0.7,-3) rectangle (2.3,6);
\draw [fill=OrangeRed!70,draw=none,xshift=19cm,yshift=-1cm] (0.7,1) rectangle (2.3,6);
\draw [fill=OrangeRed!70,draw=none,xshift=23cm,yshift=1cm] (0.7,-1) rectangle (2.3,6);
\end{scope}

\begin{scope}[xshift=3cm,yshift=3cm]
\draw [fill=NavyBlue!50,draw=none,xshift=9cm,yshift=7cm] (1,5) rectangle (18,6);
\draw [fill=Black,draw=none,xshift=9cm,yshift=21cm] (1,5) rectangle (2,6);
\draw [fill=Black,draw=none,xshift=9cm,yshift=-8cm] (1,5) rectangle (2,6);

\draw [fill=NavyBlue!50,draw=none,xshift=11cm,yshift=5cm] (1,-8) rectangle (2,6);
\draw [fill=Black,draw=none,xshift=25cm,yshift=5cm] (1,5) rectangle (2,6);
\draw [fill=Black,draw=none,xshift=-4cm,yshift=5cm] (1,5) rectangle (2,6);

\draw [fill=NavyBlue!50,draw=none,xshift=13cm,yshift=11cm] (1,5) rectangle (14,6);
\draw [fill=Black,draw=none,xshift=13cm,yshift=21cm] (1,5) rectangle (2,6);
\draw [fill=Black,draw=none,xshift=13cm,yshift=-8cm] (1,5) rectangle (2,6);

\draw [fill=NavyBlue!50,draw=none,xshift=15cm,yshift=9cm] (1,-12) rectangle (2,6);
\draw [fill=Black,draw=none,xshift=25cm,yshift=9cm] (1,5) rectangle (2,6);
\draw [fill=Black,draw=none,xshift=-4cm,yshift=9cm] (1,5) rectangle (2,6);

\draw [fill=NavyBlue!50,draw=none,xshift=17cm,yshift=15cm] (1,5) rectangle (10,6);
\draw [fill=Black,draw=none,xshift=17cm,yshift=21cm] (1,5) rectangle (2,6);
\draw [fill=Black,draw=none,xshift=17cm,yshift=-8cm] (1,5) rectangle (2,6);

\draw [fill=NavyBlue!50,draw=none,xshift=19cm,yshift=13cm] (1,-16) rectangle (2,6);
\draw [fill=Black,draw=none,xshift=25cm,yshift=13cm] (1,5) rectangle (2,6);
\draw [fill=Black,draw=none,xshift=-4cm,yshift=13cm] (1,5) rectangle (2,6);

\draw [fill=NavyBlue!50,draw=none,xshift=21cm,yshift=19cm] (1,5) rectangle (6,6);
\draw [fill=Black,draw=none,xshift=21cm,yshift=21cm] (1,5) rectangle (2,6);
\draw [fill=Black,draw=none,xshift=21cm,yshift=-8cm] (1,5) rectangle (2,6);

\draw [fill=NavyBlue!50,draw=none,xshift=23cm,yshift=17cm] (1,-20) rectangle (2,6);
\draw [fill=Black,draw=none,xshift=25cm,yshift=17cm] (1,5) rectangle (2,6);
\draw [fill=Black,draw=none,xshift=-4cm,yshift=17cm] (1,5) rectangle (2,6);

\end{scope}

\draw [fill=DodgerBlue!50,draw=none,xshift=3cm,yshift=3cm] (1,5) rectangle (2,27);
\draw [fill=DodgerBlue!50,draw=none,xshift=3cm,yshift=3cm] (3,3) rectangle (4,27);
\draw [fill=DodgerBlue!50,draw=none,xshift=3cm,yshift=3cm] (5,1) rectangle (6,27);

\begin{scope}[yshift=3cm]
\draw [fill=OrangeRed!70,draw=none,xshift=3cm,yshift=7cm] (-3,5) rectangle (2,6);
\draw [fill=OrangeRed!70,draw=none,xshift=5cm,yshift=11cm] (-5,5) rectangle (2,6);
\draw [fill=OrangeRed!70,draw=none,xshift=5cm,yshift=15cm] (-5,5) rectangle (2,6);
\draw [fill=OrangeRed!70,draw=none,xshift=7cm,yshift=19cm] (-7,5) rectangle (2,6);
\end{scope}

\begin{scope}[xshift=3cm,yshift=3cm]
\draw [fill=OrangeRed,draw=none,xshift=11cm,yshift=7cm] (1,5) rectangle (2,20);
\draw [fill=OrangeRed,draw=none,xshift=15cm,yshift=11cm] (1,5) rectangle (2,16);
\draw [fill=OrangeRed,draw=none,xshift=19cm,yshift=15cm] (1,5) rectangle (2,12);
\draw [fill=OrangeRed,draw=none,xshift=23cm,yshift=19cm] (1,5) rectangle (2,8);
\end{scope}

\draw [fill=DodgerBlue,draw=none,xshift=3cm,yshift=3cm] (1,5) rectangle (2,6);
\draw [fill=DodgerBlue,draw=none,xshift=3cm,yshift=3cm] (3,3) rectangle (4,4);
\draw [fill=DodgerBlue,draw=none,xshift=3cm,yshift=3cm] (5,1) rectangle (6,2);

\end{tikzpicture}
\end{center}
\caption{When all copies of a variable are extended upwards.}

\end{figure}
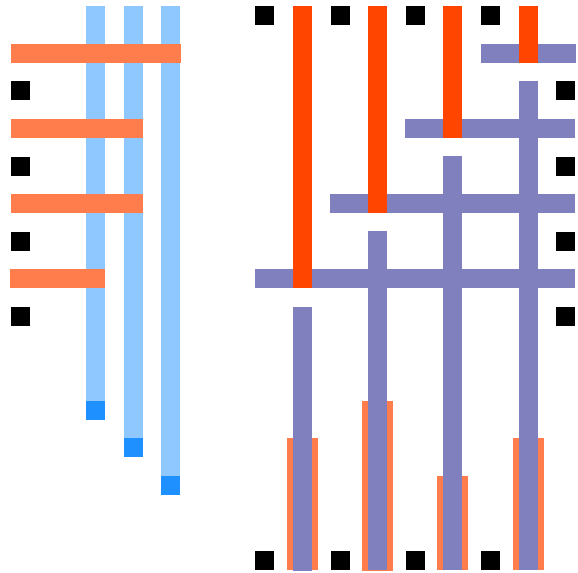

\begin{figure}[h]

\begin{center}
\begin{tikzpicture}[scale=0.25]

\draw [fill=DodgerBlue!50,draw=none,xshift=3cm,yshift=3cm] (1,5) rectangle (27,6);
\draw [fill=DodgerBlue!50,draw=none,xshift=3cm,yshift=3cm] (3,3) rectangle (27,4);
\draw [fill=DodgerBlue!50,draw=none,xshift=3cm,yshift=3cm] (5,1) rectangle (27,2);

\begin{scope}[xshift=3cm]
\draw [fill=OrangeRed!70,draw=none,xshift=11cm,yshift=1cm] (1,-1) rectangle (2,6);
\draw [fill=OrangeRed!70,draw=none,xshift=15cm,yshift=3cm] (1,-3) rectangle (2,6);
\draw [fill=OrangeRed!70,draw=none,xshift=19cm,yshift=-1cm] (1,1) rectangle (2,6);
\draw [fill=OrangeRed!70,draw=none,xshift=23cm,yshift=1cm] (1,-1) rectangle (2,6);
\end{scope}

\begin{scope}[yshift=3cm]
\draw [fill=OrangeRed!70,draw=none,xshift=3cm,yshift=7cm] (-3,4.7) rectangle (2,6.3);
\draw [fill=OrangeRed!70,draw=none,xshift=5cm,yshift=11cm] (-5,4.7) rectangle (2,6.3);
\draw [fill=OrangeRed!70,draw=none,xshift=5cm,yshift=15cm] (-5,4.7) rectangle (2,6.3);
\draw [fill=OrangeRed!70,draw=none,xshift=7cm,yshift=19cm] (-7,4.7) rectangle (2,6.3);
\end{scope}

\begin{scope}[xshift=3cm,yshift=3cm]
\draw [fill=NavyBlue!50,draw=none,xshift=9cm,yshift=7cm] (-12,5) rectangle (2,6);
\draw [fill=Black,draw=none,xshift=9cm,yshift=21cm] (1,5) rectangle (2,6);
\draw [fill=Black,draw=none,xshift=9cm,yshift=-8cm] (1,5) rectangle (2,6);

\draw [fill=NavyBlue!50,draw=none,xshift=11cm,yshift=5cm] (1,5) rectangle (2,22);
\draw [fill=Black,draw=none,xshift=25cm,yshift=5cm] (1,5) rectangle (2,6);
\draw [fill=Black,draw=none,xshift=-4cm,yshift=5cm] (1,5) rectangle (2,6);

\draw [fill=NavyBlue!50,draw=none,xshift=13cm,yshift=11cm] (-16,5) rectangle (2,6);
\draw [fill=Black,draw=none,xshift=13cm,yshift=21cm] (1,5) rectangle (2,6);
\draw [fill=Black,draw=none,xshift=13cm,yshift=-8cm] (1,5) rectangle (2,6);

\draw [fill=NavyBlue!50,draw=none,xshift=15cm,yshift=9cm] (1,5) rectangle (2,18);
\draw [fill=Black,draw=none,xshift=25cm,yshift=9cm] (1,5) rectangle (2,6);
\draw [fill=Black,draw=none,xshift=-4cm,yshift=9cm] (1,5) rectangle (2,6);

\draw [fill=NavyBlue!50,draw=none,xshift=17cm,yshift=15cm] (-20,5) rectangle (2,6);
\draw [fill=Black,draw=none,xshift=17cm,yshift=21cm] (1,5) rectangle (2,6);
\draw [fill=Black,draw=none,xshift=17cm,yshift=-8cm] (1,5) rectangle (2,6);

\draw [fill=NavyBlue!50,draw=none,xshift=19cm,yshift=13cm] (1,5) rectangle (2,14);
\draw [fill=Black,draw=none,xshift=25cm,yshift=13cm] (1,5) rectangle (2,6);
\draw [fill=Black,draw=none,xshift=-4cm,yshift=13cm] (1,5) rectangle (2,6);

\draw [fill=NavyBlue!50,draw=none,xshift=21cm,yshift=19cm] (-24,5) rectangle (2,6);
\draw [fill=Black,draw=none,xshift=21cm,yshift=21cm] (1,5) rectangle (2,6);
\draw [fill=Black,draw=none,xshift=21cm,yshift=-8cm] (1,5) rectangle (2,6);

\draw [fill=NavyBlue!50,draw=none,xshift=23cm,yshift=17cm] (1,5) rectangle (2,10);
\draw [fill=Black,draw=none,xshift=25cm,yshift=17cm] (1,5) rectangle (2,6);
\draw [fill=Black,draw=none,xshift=-4cm,yshift=17cm] (1,5) rectangle (2,6);

\end{scope}

\begin{scope}[xshift=3cm,yshift=3cm]
\draw [fill=OrangeRed,draw=none,xshift=11cm,yshift=7cm] (1,5) rectangle (16,6);
\draw [fill=OrangeRed,draw=none,xshift=15cm,yshift=11cm] (1,5) rectangle (12,6);
\draw [fill=OrangeRed,draw=none,xshift=19cm,yshift=15cm] (1,5) rectangle (8,6);
\draw [fill=OrangeRed,draw=none,xshift=23cm,yshift=19cm] (1,5) rectangle (4,6);
\end{scope}

\draw [fill=DodgerBlue,draw=none,xshift=3cm,yshift=3cm] (1,5) rectangle (2,6);
\draw [fill=DodgerBlue,draw=none,xshift=3cm,yshift=3cm] (3,3) rectangle (4,4);
\draw [fill=DodgerBlue,draw=none,xshift=3cm,yshift=3cm] (5,1) rectangle (6,2);

\end{tikzpicture}
\end{center}

\caption{When all copies of a variable are extended to the right.}
\end{figure}
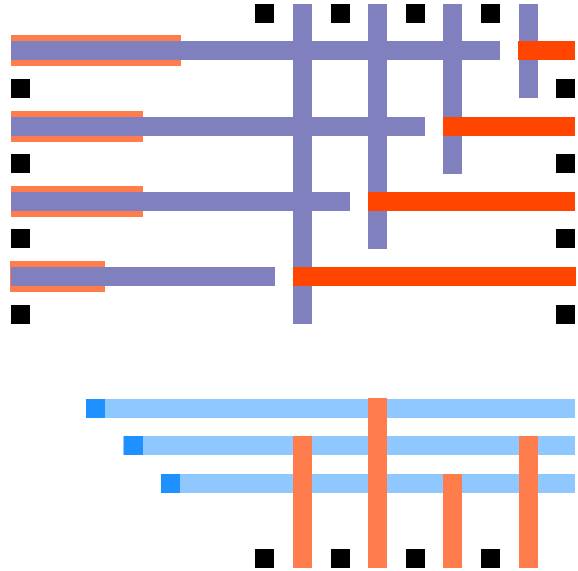

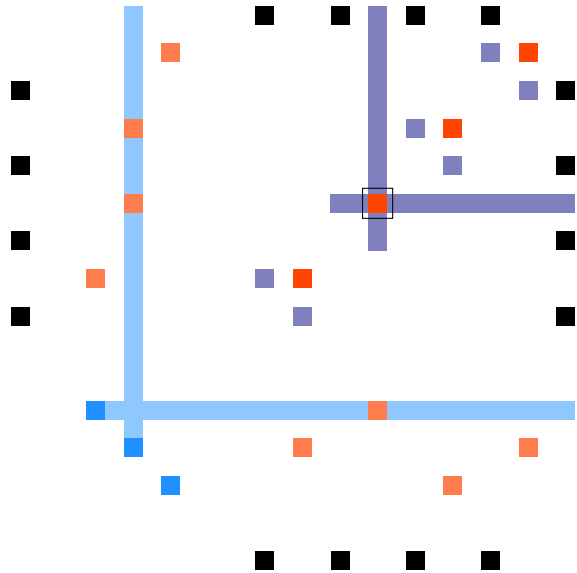
\begin{figure}[h]

\begin{center}
\begin{tikzpicture}[scale=0.25]

\draw [fill=DodgerBlue!50,draw=none,xshift=3cm,yshift=3cm] (1,5) rectangle (27,6);
\draw [fill=DodgerBlue!50,draw=none,xshift=3cm,yshift=3cm] (3,3) rectangle (4,27);

\draw [fill=DodgerBlue,draw=none,xshift=3cm,yshift=3cm] (1,5) rectangle (2,6);
\draw [fill=DodgerBlue,draw=none,xshift=3cm,yshift=3cm] (3,3) rectangle (4,4);
\draw [fill=DodgerBlue,draw=none,xshift=3cm,yshift=3cm] (5,1) rectangle (6,2);

\begin{scope}[yshift=3cm]
\draw [fill=OrangeRed!70,draw=none,xshift=3cm,yshift=7cm] (1,5) rectangle (2,6);
\draw [fill=OrangeRed!70,draw=none,xshift=5cm,yshift=11cm] (1,5) rectangle (2,6);
\draw [fill=OrangeRed!70,draw=none,xshift=5cm,yshift=15cm] (1,5) rectangle (2,6);
\draw [fill=OrangeRed!70,draw=none,xshift=7cm,yshift=19cm] (1,5) rectangle (2,6);
\end{scope}

\begin{scope}[xshift=3cm]
\draw [fill=OrangeRed!70,draw=none,xshift=11cm,yshift=1cm] (1,5) rectangle (2,6);
\draw [fill=OrangeRed!70,draw=none,xshift=15cm,yshift=3cm] (1,5) rectangle (2,6);
\draw [fill=OrangeRed!70,draw=none,xshift=19cm,yshift=-1cm] (1,5) rectangle (2,6);
\draw [fill=OrangeRed!70,draw=none,xshift=23cm,yshift=1cm] (1,5) rectangle (2,6);
\end{scope}

\begin{scope}[xshift=3cm,yshift=3cm]
\draw [fill=NavyBlue!50,draw=none,xshift=9cm,yshift=7cm] (1,5) rectangle (2,6);
\draw [fill=Black,draw=none,xshift=9cm,yshift=21cm] (1,5) rectangle (2,6);
\draw [fill=Black,draw=none,xshift=9cm,yshift=-8cm] (1,5) rectangle (2,6);

\draw [fill=NavyBlue!50,draw=none,xshift=11cm,yshift=5cm] (1,5) rectangle (2,6);
\draw [fill=Black,draw=none,xshift=25cm,yshift=5cm] (1,5) rectangle (2,6);
\draw [fill=Black,draw=none,xshift=-4cm,yshift=5cm] (1,5) rectangle (2,6);

\draw [fill=NavyBlue!50,draw=none,xshift=13cm,yshift=11cm] (1,5) rectangle (14,6);
\draw [fill=Black,draw=none,xshift=13cm,yshift=21cm] (1,5) rectangle (2,6);
\draw [fill=Black,draw=none,xshift=13cm,yshift=-8cm] (1,5) rectangle (2,6);

\draw [fill=NavyBlue!50,draw=none,xshift=15cm,yshift=9cm] (1,5) rectangle (2,18);
\draw [fill=Black,draw=none,xshift=25cm,yshift=9cm] (1,5) rectangle (2,6);
\draw [fill=Black,draw=none,xshift=-4cm,yshift=9cm] (1,5) rectangle (2,6);

\draw [fill=NavyBlue!50,draw=none,xshift=17cm,yshift=15cm] (1,5) rectangle (2,6);
\draw [fill=Black,draw=none,xshift=17cm,yshift=21cm] (1,5) rectangle (2,6);
\draw [fill=Black,draw=none,xshift=17cm,yshift=-8cm] (1,5) rectangle (2,6);

\draw [fill=NavyBlue!50,draw=none,xshift=19cm,yshift=13cm] (1,5) rectangle (2,6);
\draw [fill=Black,draw=none,xshift=25cm,yshift=13cm] (1,5) rectangle (2,6);
\draw [fill=Black,draw=none,xshift=-4cm,yshift=13cm] (1,5) rectangle (2,6);

\draw [fill=NavyBlue!50,draw=none,xshift=21cm,yshift=19cm] (1,5) rectangle (2,6);
\draw [fill=Black,draw=none,xshift=21cm,yshift=21cm] (1,5) rectangle (2,6);
\draw [fill=Black,draw=none,xshift=21cm,yshift=-8cm] (1,5) rectangle (2,6);

\draw [fill=NavyBlue!50,draw=none,xshift=23cm,yshift=17cm] (1,5) rectangle (2,6);
\draw [fill=Black,draw=none,xshift=25cm,yshift=17cm] (1,5) rectangle (2,6);
\draw [fill=Black,draw=none,xshift=-4cm,yshift=17cm] (1,5) rectangle (2,6);

\end{scope}

\begin{scope}[xshift=3cm,yshift=3cm]
\draw [fill=OrangeRed,draw=none,xshift=11cm,yshift=7cm] (1,5) rectangle (2,6);
\draw [fill=OrangeRed,draw=none,xshift=15cm,yshift=11cm] (1,5) rectangle (2,6);
\draw [fill=none,draw=black,xshift=15cm,yshift=11cm] (0.7,4.7) rectangle (2.3,6.3);
\draw [fill=OrangeRed,draw=none,xshift=19cm,yshift=15cm] (1,5) rectangle (2,6);
\draw [fill=OrangeRed,draw=none,xshift=23cm,yshift=19cm] (1,5) rectangle (2,6);
\end{scope}
\end{tikzpicture}
\end{center}

\caption{A violation when two copies are extended in different directions.}

\end{figure}

\section{Future Directions}

Studying this natural optimization version of \RE{} leads us to several new questions. First, to obtain a constant-factor approximation algorithm that is independent of $d$, we would like to be able to answer the question of whether at least $k$ rectangles can be pushed along one direction in polynomial time, and further address the question of whether $k$ rectangles can be pushed up or down with density at most $d$ in polynomial time. From the reduction in~\cite{MKWY11} it can be seen that the question of whether all rectangles can be pushed with density at most three when the only available directions are top and right, is already \NPH{}. It would be interesting to examine what happens when the combinations of directions that are available are parallel (like up and down, or right and left), and one of the motivations is that this directly impacts the approximation ratio. 


There are unresolved questions in the parameterized context as well. For example, is the problem fixed-parameter tractable when the input configuration has points of density $d$? Further, for the cases when the input configuration has density at most $(d-1)$, the algorithm presented here has a running time of $2^{O(k \log k)}n^{O(1)}$. Can this be improved, for instance, $2^{O(k)} n^{O(1)}$? 

A general direction of interest is to obtain substantially improved algorithms for the special case when the rectangles are squares aligned to a grid, for which we establish NP-hardness here. 

\section*{Acknowledgements} We would like to thank Sue Whitesides who suggested to consider the problem in $3$-dimensions.
%
%

\bibliographystyle{abbrv}
\bibliography{RE}

\begin{thebibliography}{10}

\bibitem{Ahlswede06}
R.~Ahlswede and I.~Karapetyan.
\newblock Intersection graphs of rectangles and segments.
\newblock {\em General Theory of Information Transfer and Combinatorics},
  4123:1064--1065, 2006.

\bibitem{AD09}
B.~M. Anthony and R.~Denman.
\newblock {k-Bounded Positive Not All Equal LE3SAT}.
\newblock In {\em Brown Working Papers}, 2009.

\bibitem{Aschner2013}
R.~Aschner, M.~Katz, G.~Morgenstern, and Y.~Yuditsky.
\newblock Approximation schemes for covering and packing.
\newblock {\em WALCOM: Algorithms and Computation}, 7748:89--100, 2013.

\bibitem{Asplund1960}
E.~Asplund and B.~Gr\"{u}nbaum.
\newblock {On a coloring problem}.
\newblock {\em Math. Scand.}, 8:181--188, 1960.

\bibitem{AEYZ13}
S.~Assadi, E.~Emamjomeh-Zadeh, S.~Yazdanbod, and H.~Zarrabi-Zadeh.
\newblock On the rectangle escape problem.
\newblock In {\em Canadian Conference on Computational Geometry (CCCG)}, pages
  235--240, 2013.

\bibitem{DBLP:conf/soda/ChalermsookC09}
P.~Chalermsook and J.~Chuzhoy.
\newblock Maximum independent set of rectangles.
\newblock In {\em SODA}, pages 892--901, 2009.

\bibitem{Chan2011}
T.~M. Chan and S.~Har-Peled.
\newblock {Approximation Algorithms for Maximum Independent Set of
  Pseudo-Disks}.
\newblock {\em Discrete and Computational Geometry}, pages 373--392, 2012.

\bibitem{DBLP:dblp_conf/latin/CorreaFS14}
J.~Correa, L.~Feuilloley, P.~P\'{e}rez-Lantero, and J.~A. Soto.
\newblock {Independent and hitting sets of rectangles intersecting a diagonal
  line: Algorithms and complexity}.
\newblock {\em Discrete \& Computational Geometry}, 53(2):344--365, 2015.

\bibitem{downey1999parameterized}
R.~G. Downey and M.~R. Fellows.
\newblock {\em Parameterized complexity}, volume~3.
\newblock Springer Heidelberg, 1999.

\bibitem{Ene2011a}
A.~Ene, S.~Har-Peled, and B.~Raichel.
\newblock {Geometric Packing under Non-uniform Constraints}.
\newblock {\em CoRR}, abs/1107.2, 2011.

\bibitem{faigle1995note}
U.~Faigle and W.~M. Nawijn.
\newblock Note on scheduling intervals on-line.
\newblock {\em Discrete Applied Mathematics}, 58(1):13--17, 1995.

\bibitem{Fellows200953}
M.~R. Fellows, D.~Hermelin, F.~Rosamond, and S.~Vialette.
\newblock On the parameterized complexity of multiple-interval graph problems.
\newblock {\em Theoretical Computer Science}, 410(1):53--61, 2009.

\bibitem{flum2006parameterized}
J.~Flum and M.~Grohe.
\newblock {\em Parameterized {C}omplexity {T}heory}.
\newblock Springer, 2006.

\bibitem{DBLP:journals/dm/GyarfasL85}
A.~Gy{\'a}rf{\'a}s and J.~Lehel.
\newblock Covering and coloring problems for relatives of intervals.
\newblock {\em Discrete Mathematics}, 55(2):167--180, 1985.

\bibitem{KMYW10}
H.~Kong, Q.~Ma, T.~Yan, and M.~D.~F. Wong.
\newblock An optimal algorithm for finding disjoint rectangles and its
  application to pcb routing.
\newblock In {\em Proceedings of the 47th Design Automation Conference}, DAC
  '10, pages 212--217, New York, NY, USA, 2010. ACM.

\bibitem{MKWY11}
Q.~Ma, H.~Kong, M.~D. Wong, and E.~F. Young.
\newblock A provably good approximation algorithm for rectangle escape problem
  with application to pcb routing.
\newblock In {\em Proceedings of the 16th Asia and South Pacific Design
  Automation Conference}, pages 843--848. IEEE Press, 2011.

\bibitem{niedermeier2006invitation}
R.~Niedermeier.
\newblock {\em Invitation to Fixed-Parameter Algorithms}.
\newblock Oxford University Press, 2006.

\bibitem{Palios:1997:CMN:249088.249092}
L.~Palios.
\newblock Connecting the maximum number of nodes in the grid to the boundary
  with nonintersecting line segments.
\newblock {\em J. Algorithms}, 22(1):57--92, Jan. 1997.

\bibitem{DBLP:conf/cccg/RoyGMS14}
A.~B. Roy, S.~Govindarajan, N.~Misra, and S.~Shetty.
\newblock On the d-runaway rectangle escape problem.
\newblock In {\em Proceedings of the 26th Canadian Conference on Computational
  Geometry, {CCCG} 2014, Halifax, Nova Scotia, Canada, 2014}, 2014.

\bibitem{toth2004handbook}
C.~D. Toth, J.~O'Rourke, and J.~E. Goodman.
\newblock {\em Handbook of discrete and computational geometry}.
\newblock CRC press, 2004.

\end{thebibliography}

\end{document}